\documentclass[aps,twocolumn,prl,superscriptaddress,amsmath,amssymb,amsfonts]{revtex4-1}

\usepackage[T1]{fontenc}
\usepackage[latin9]{inputenc}
\usepackage{lmodern} % load a font with all the characters
\usepackage{color}
\usepackage{bbold}
\usepackage{dsfont}
\usepackage{graphicx}
\usepackage[caption=false]{subfig}
\usepackage[colorlinks]{hyperref}
\begin{document}

\global\long\def\eqn#1{\begin{align}#1\end{align}}
\global\long\def\vec#1{\overrightarrow{#1}}
\global\long\def\ket#1{\left|#1\right\rangle }
\global\long\def\bra#1{\left\langle #1\right|}
\global\long\def\bkt#1{\left(#1\right)}
\global\long\def\sbkt#1{\left[#1\right]}
\global\long\def\cbkt#1{\left\{#1\right\}}
\global\long\def\abs#1{\left\vert#1\right\vert}
\global\long\def\cev#1{\overleftarrow{#1}}
\global\long\def\der#1#2{\frac{{d}#1}{{d}#2}}
\global\long\def\pard#1#2{\frac{{\partial}#1}{{\partial}#2}}
\global\long\def\re{\mathrm{Re}}
\global\long\def\im{\mathrm{Im}}
\global\long\def\dd{\mathrm{d}}

\global\long\def\avg#1{\left\langle #1 \right\rangle}
\global\long\def\mr#1{\mathrm{#1}}
\global\long\def\mb#1{{\mathbf #1}}
\global\long\def\mc#1{\mathcal{#1}}
\global\long\def\tr{\mathrm{Tr}}
\global\long\def\dbar#1{\Bar{\Bar{#1}}}

\global\long\def\nth{$n^{\mathrm{th}}$\,}
\global\long\def\mth{$m^{\mathrm{th}}$\,}
\global\long\def\non{\nonumber}

\newcommand{\orange}[1]{{\color{orange} {#1}}}
\newcommand{\cyan}[1]{{\color{cyan} {#1}}}
\newcommand{\blue}[1]{{\color{blue} {#1}}}
\newcommand{\yellow}[1]{{\color{yellow} {#1}}}
\newcommand{\green}[1]{{\color{green} {#1}}}
\newcommand{\red}[1]{{\color{red} {#1}}}
\newcommand{\slr}[1]{{\color{purple} {#1}}}
\global\long\def\todo#1{\orange{{$\bigstar$ \cyan{\bf\sc #1}}$\bigstar$} }
\global\long\def\addref{\orange{{$\bigstar$ \blue{\bf\sc Add Ref}}$\bigstar$} }

\title{Observation of vacuum-induced collective quantum beats}

\author{Hyok Sang Han}
\affiliation{Joint Quantum Institute, University of Maryland and the National Institute of Standards and Technology, College Park, Maryland 20742, USA}
\author{Ahreum Lee}
\affiliation{Joint Quantum Institute, University of Maryland and the National Institute of Standards and Technology, College Park, Maryland 20742, USA}

\author{Kanupriya Sinha} 
\email{kanu@princeton.edu}
\affiliation{Department of Electrical Engineering, Princeton University, Princeton, New Jersey 08544, USA}

\author{Fredrik K. Fatemi}
\affiliation{U.S. Army Research Laboratory, Adelphi, Maryland 20783, USA}
\affiliation{Quantum Technology Center, University of Maryland, College Park, MD 20742, USA}
\author{S. L. Rolston}
\email{rolston@umd.edu}
\affiliation{Joint Quantum Institute, University of Maryland and the National Institute of Standards and Technology, College Park, Maryland 20742, USA}
\affiliation{Quantum Technology Center, University of Maryland, College Park, MD 20742, USA}

\begin{abstract}
	
	We demonstrate collectively enhanced vacuum-induced quantum beat dynamics from a three-level V-type atomic system. Exciting a dilute atomic gas of magneto-optically trapped $^{85}$Rb atoms with a weak drive resonant on one of the transitions, we observe the forward-scattered field after a sudden shut-off of the laser. The subsequent radiative dynamics, measured for various optical depths of the atomic cloud, exhibits superradiant decay rates, as well as collectively enhanced  quantum beats. Our work is also the first experimental illustration of quantum beats  arising from atoms initially prepared in a single excited level as a result of the vacuum-induced coupling between excited levels.

\end{abstract}

\maketitle

\textit{Introduction.}---Quantum beats are a well-studied phenomenon that describes the interference between spontaneously emitted radiation from two or more excited levels, resulting in a periodic modulation of the radiated field intensity \cite{Jaynes_1980}. This has been a valuable spectroscopic tool to measure the energy difference between excited levels across many experimental platforms such as  atoms \cite{Haroche_1973, wade_2014}, molecules \cite{Hack_1991}, semiconductors \cite{Stolz_1991}, and quantum dots \cite{Kozin_2002, Bylsma_2012}. 

Although quantum beats have been extensively studied, here we demonstrate two new aspects: (i) quantum beats without an initial superposition of excited levels, and (ii) enhanced beat amplitudes due to collective emission of light \cite{Dicke_1954, Gross_1982}. In a typical quantum beat experiment, an excitation pulse with sufficient bandwidth to span the energy spacing between multiple excited atomic levels is used to create an initial coherent superposition. The beat signal amplitude is proportional to the coherence between the excited levels, and in the absence of an initial superposition, one might expect no quantum beats. This notion was challenged in \cite{Hegerfeldt_1993, Hegerfeldt_1994}, predicting that the vacuum electromagnetic (EM) field can create the required coherence between the excited atomic levels. However, experimental observation of such vacuum-induced quantum beats is challenging due to the competing requirements on the level structure: The excited levels separation needs to be large compared to the natural linewidth to enable the initialization of only one of the levels, which, in turn, reduces the strength of the vacuum-induced coupling. 

We experimentally address this using the well-separated $^{85}$Rb $^{5}P_{3/2}$ $F'=3$ and 4 hyperfine levels as our excited levels and using a long enough (200 ns) excitation pulse such that any coherence due to the turn-on edge decays away, leaving the atomic population in a single excited level. Detecting the forward-scattered mode (see Fig.\,\ref{fig_schematic}\,(a)) allows us to observe the radiation from a timed-Dicke state \cite{Scully_2006, Bienaime_2013, Bromley_Ye_2016}. We theoretically illustrate that for such a collective state, the quantum beat dynamics can be cooperatively enhanced by the constructive interference between the transition processes in different atoms. The collective amplification of the forward-scattered beat signal allows us to observe vacuum-induced quantum beats and serves as an experimental proof of collective effects in quantum beats. Such collective enhancement may also be used to amplify small signals that are otherwise unobservable.

\textit{Model.}---Let us consider a system of  three-level V-type $^{85}$Rb atoms, with the ground level  $\ket{1}=\ket{5S_{1/2}, F=3}$ and the two excited levels  $\ket{2}=\ket{5P_{3/2},F'=4}$ and $\ket{3}=\ket{5P_{3/2},F'=3}$ (see Fig.\,\ref{fig_schematic}\,(b)). The frequency difference between the excited levels is $\omega_{23}=2\pi\cdot 121$ MHz, and the optical transition wavelength between the ground and the excited levels is $\lambda = 780$ nm. We observe the forward scattering, where the phase factor of the field from propagation within the atomic cloud is exactly compensated by the phases of the atomic dipoles initially induced by the drive \cite{Scully_2006}. The damping rates of atomic levels originating from second-order coupling between $\ket{j}$ and $\ket{l}$ is $
\Gamma_{jl} = \frac{\vec{d}_{j1} \cdot \vec{d}_{l1}\omega_{j1}^3}{3\pi \varepsilon_0 \hbar c^3}$, where $\vec{d}_{j1}$ and $\omega_{j1}$ are the transition dipole moments and the transition frequency between $\ket{j}$ and $\ket{1}$, respectively. Note that $\Gamma_{23}$ represents second-order coupling between the excited states via vacuum-induced decay and absorption \cite{Hegerfeldt_1994}, while $\Gamma_{22}$ and $\Gamma_{33}$ describe the normal decay of the excited states.  Assuming that all the transition dipole moments are real and parallel to each other  $\Gamma_{23}\approx\sqrt{\Gamma_{22}\Gamma_{33}}$. In our system $\Gamma_{22}=2\pi\cdot6.1$ MHz is the single-atom decay rate of the $5P_{3/2}$ level and $\Gamma_{33}=\frac{5}{9}\Gamma_{22}$, as $\ket{3}$ decays to $\ket{1}$ only fractionally with the branching ratio 5/9 \cite{Steck_Rb85}.

The atoms are initialized in a symmetric state with a shared \emph{single} excitation in $ \ket{2}$. After a sudden turn-off of the drive field, the atomic ensemble starts to decay due to its interaction with the vacuum field modes, which couple the excited levels to reveal quantum beating. Analytically solving collective atomic and field dynamics in
the experimental regime where the excited atomic levels
are well-separated from each other ($\Gamma_{jl}^{(N)}\ll\omega_{23}$), we find the intensity of light emitted from the ensemble as (see Supplemental Material)
\eqn{
	\frac{I(t)}{I_0} = e^{-\Gamma^{(N)}_{22} t}+I_\mathrm{b} e^{-\Gamma^{(N)}_\mathrm{avg} t} \sin{\left(\omega_{23} t+\phi \right)},
	\label{eq_intensity}
}
where we have defined the total collective  decay rate as $ \Gamma_{jl}^{(N)}\equiv (1 + Nf )\Gamma_{jl}$, with  $f$ corresponding to the angular emission factor in to the forward scattered modes and $N$ corresponding to the effective number of atoms emitting collectively \cite{Araujo_2016}. We have assumed here that the atoms emit collectively in the forward direction as a result of the phase coherence due to the timed-Dicke state, while the emission in the remainder of the modes is independent. $\Gamma_{\text{avg}}^{(N)}\equiv \bkt{\Gamma_{22}^{(N)} + \Gamma_{33}^{(N)}}/2$ is the average decay rate of excited levels, and the relative beat intensity is defined as
\eqn{
	I_\mathrm{b} = \frac{\Gamma_{33}^{(N)}}{\omega_{23}}\approx \frac{5}{9}\frac{\Gamma_{22}^{(N)}}{\omega_{23}},
	\label{eq_beat_amp}
}
and the beat phase is defined as
\eqn{
	\phi = \arctan\bkt{\frac{\Gamma^{(N)}_{22}}{\omega_{23}}}.
	\label{eq_phase}
}
The first term of Eq.\,\eqref{eq_intensity} represents the collective decay from $\ket{2}$, with a cooperatively enhanced amplitude and decay rate  relative to a single atom. The second term accounts for the small but non-negligible beat which decays away with an enhanced average rate $ \Gamma_\mr{avg}^{(N)}$. This result shows that vacuum-induced quantum beats in the absence of an initial superposition of excited atomic levels can exhibit collective effects, generalizing the single atom quantum trajectory prediction in \cite{Hegerfeldt_1994}. From Eq.\,\eqref{eq_beat_amp} we observe that the collective nature of the quantum beat originates from the virtual coupling between the excited levels as indicated by the cross-damping term $\Gamma_{23}$.

\begin{figure}[]
	\centering
	\includegraphics[width = 3.5 in]{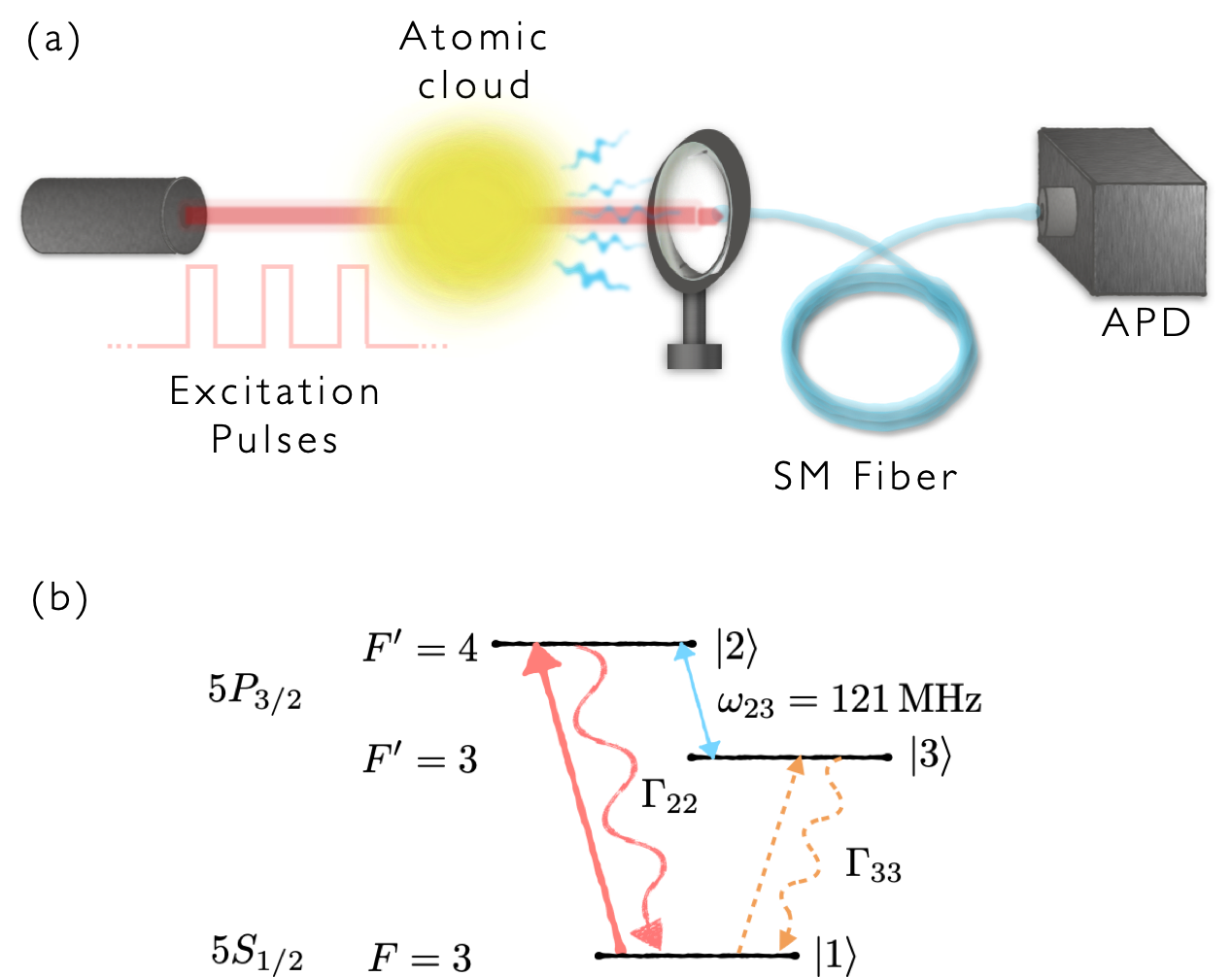}
	\caption{\textbf{(a)} Experimental setup. A linearly polarized excitation beam containing a train of pulses illuminates a cold $^{85}$Rb atomic cloud produced by the MOT. The photons scattered by the cloud in the forward direction are coupled into the single-mode (SM) fiber, counted by the avalanche photodiode (APD) and histogrammed to obtain the atomic radiative decay profile. \textbf{(b)} Relevant energy levels of $^{85}$Rb atom. The excitation beam (780 nm) resonantly drives the $\ket{1}$$\leftrightarrow$$\ket{2}$ transition. $\Gamma_{22}$ and $\Gamma_{33}$ are the decay rates of the excited levels $\ket{2}$ and $\ket{3}$, respectively, to the ground level $\ket{1}$.}
	\label{fig_schematic}
\end{figure}

\begin{figure*}[t]
	\centering
	\includegraphics[width = 6.5 in]{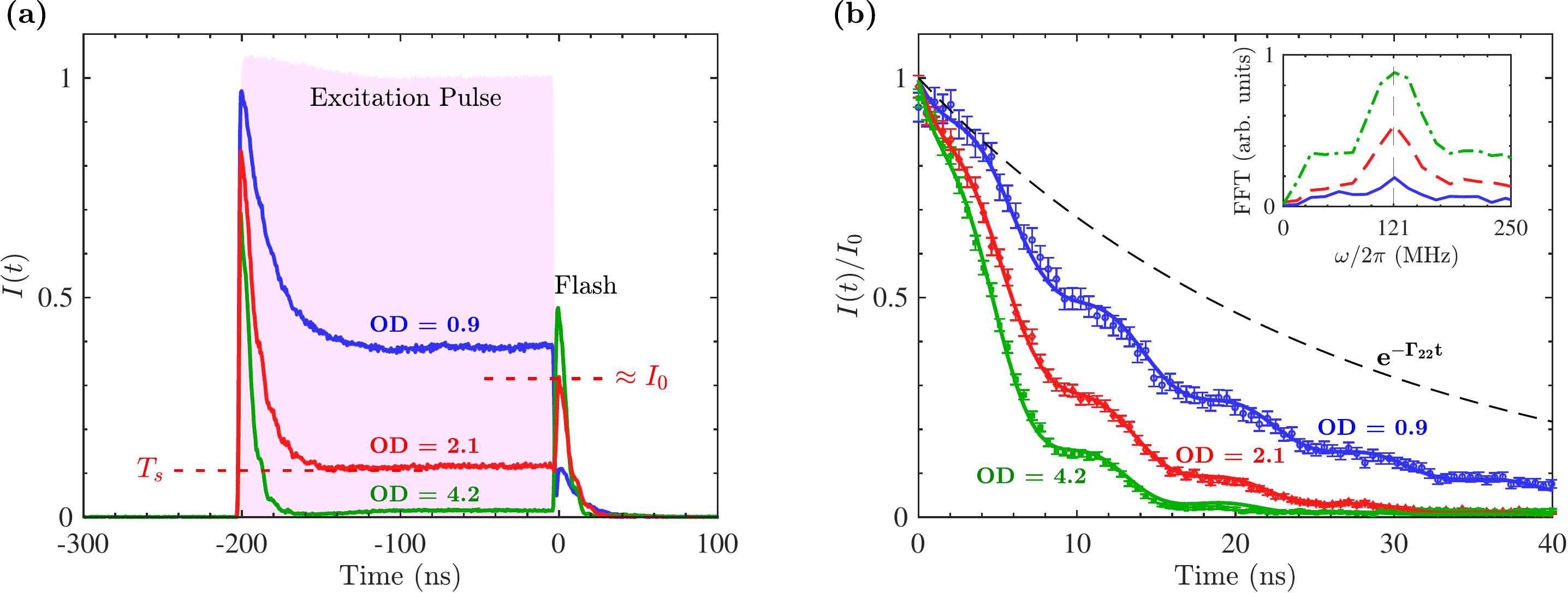}
	\caption{\textbf{(a)} Examples of the histogrammed photon counts for various optical depths (OD), representing the forward-mode intensity, normalized to that of the excitation pulse. As the excitation pulse is abruptly (within $\approx$ 3.5 ns) turned off, the flash of the photon emission occurs with the peak intensity $I_0$ proportional to the OD. \textbf{(b)} The decay profiles after the flash peak are zoomed-in for analysis. The intensity of each curve is further normalized to the exponential decay amplitude $I_0$ ($\approx$ flash peak size shown in (a)). The error bars represent the shot-noise limit of the photon counts. The overlaid solid lines fit the data using Eq.\,\eqref{eq_intensity} whose results are displayed in Fig.\,\ref{fig_phase} in detail. The black dashed line represents the single-atom decay curve $I(t)=e^{-\Gamma_{22}t}$ ($\Gamma_{22}=2\pi\cdot 6.1$ MHz) without considering  collective effects. In the inset, the gray shaded region represents the absolute value of the fast Fourier transform (FFT) of the beat signal for OD = 0.9 (solid-blue), 2.1 (dashed-red), and 4.2 (dash-dot green line) showing a peak at the splitting between the two excited levels (see Supplemental Material for detail).}
	\label{fig_decay}
\end{figure*}

\textit{Experiment.}---Fig.\,\ref{fig_schematic}\,(a) shows the schematic of the experiment. A cold atomic cloud of ${\sim}10^8~^{85}$Rb atoms is produced by a magneto-optical trap (MOT) with Gaussian-shaped atomic density distribution having a $1/e$ diameter of ${\sim}2$\,mm. The ensemble satisfies the dilute regime, $\rho\lambda^3\ll 1$, where $\rho$ is the spatial atomic density, meaning that the separation between atoms is much larger than the photon wavelength. An excitation beam with $1/e^2$ diameter of 1.6 mm is overlapped with the cloud whose transmitted light is collected by a single-mode (SM) fiber 0.6-meter away in the forward direction. 

For the observation of the spontaneous emission, the MOT lasers are turned off for 200 $\mu$s during which atoms initialized in $\ket{1}$ are illuminated by a train of excitation pulses that resonantly drive the $\ket{1}$$\leftrightarrow$$\ket{2}$ transition. The peak intensity of the excitation beam is ${\sim}6\times10^{8}$ times smaller than the saturation intensity $I_\mathrm{s} = $~3.9 mW/cm$^2$ of the transition \cite{Steck_Rb85}, delivering less than one  photon per pulse on average, ensuring that the system is well within the single-excitation regime. Each excitation pulse is turned on (off) for 200 ns (800 ns)  with $>$30 dB extinction and a 3.5-ns fall-time controlled by two fibered Mach-Zehnder intensity modulators (EOSPACE AZ-0K5-10-PFA-PFA-780) in series. We derive the optical Bloch equations for the atoms in the presence of the drive and solve those numerically to obtain estimates for the population in level $\ket{3}$ (see Supplemental Material). In the  steady state at the end of the excitation pulse, the atomic ensemble is mostly in the ground state, with a small population of ${\sim}10^{-10}$  in $\ket{2}$. The population in $\ket{3}$, and the coherence between level $\ket{2}$ and $\ket{3}$  is negligibly small. Modeling the 3.5 ns laser turn off edge as a cosine-fourth function, we find that it generates negligible amplitude in $\ket{3}$ due to the small Rabi frequency and short evolution time.

After the driving field is switched off, spontaneously emitted photons coupled to the SM fiber are counted by an avalanche photodiode (APD) and histogrammed with 0.5-ns resolution. By detecting only those photons coupled to the SM fiber, we effectively filter out incoherent fluorescence, owing to the small collection solid angle ($\approx 6\times10^{-6}$ sr). The atomic velocity $v_\mathrm{} \approx 120$ nm/$\mu$s corresponding to the Doppler temperature $T_\mathrm{D}\approx150$ $\mu$K gives negligible motion compared to the optical wavelength (780 nm) within the time scale of the emission process ($1/\Gamma^{(N)}_{22}\leq$ 26 ns). After the repetition of 200 pulses  within 200-$\mu$s, the MOT lasers are turned back on to recover and maintain the atomic cloud for 1.8 ms before a new measurement cycle begins, repeating the whole sequence every 2 ms. For typical histogrammed data, we run the sequence continuously for 30 minutes, comprising $2\times10^8$ excitation pulses.

Examples of histogrammed photon counts are shown in Fig.\,\ref{fig_decay}\,(a) where $I(t)$ represents the intensity of the forward-scattered light normalized to the steady-state intensity of the excitation pulse. The atomic samples are almost transparent at the sharp switch-on edge of the excitation pulse due to its broad spectral components, but the transmission soon decays to a steady-state value $T_\mathrm{s}$, which we use to calculate the optical depth (OD $=-\ln{T_\mathrm{s}}$). We vary the OD of the MOT cloud between 0.5 and 4.5 by adjusting the injection current running through the rubidium dispensers (SAES Getters RB/NF/7/25) between 3.5 A and 6.5 A to increase atomic background pressure. The steady state transmission $T_\mathrm{s}$ results from the destructive interference between the driving field and the field coherently radiated (with  $\pi$-phase shift) in the forward direction by the atomic dipoles. When the driving field is switched off, only the atomic radiation field remains in the forward direction, resulting in a sudden intensity jump (``flash''), which has been intensively investigated in recent studies \cite{Chalony_2011, Kwong_2014, Kwong_2015}. The flash peak intensity, which is proportional to the OD, represents the intensity $I_0$ of the overall decay as in Eq.\,\eqref{eq_beat_amp}.

The decay profiles after the flash peak are magnified in Fig.\,\ref{fig_decay}\,(b) for detailed analysis. Each curve is normalized to the exponential decay amplitude $I_0$ (see Eq.\,\eqref{eq_intensity}), so the enhanced decay rates and the relative beat intensities for different OD can be easily compared. For comparision, the single-atom decay curve $I(t)=e^{-\Gamma_{22}t}$ with no collective enhancement is also shown (black dashed line). We first note that a higher OD results in an enhanced decay rate demonstrating the collective nature of the emission process. The quantum beat signal is apparent as a sinusoidal modulation of the exponential decay. This illustrates the occurrence of quantum beats in the absence of an initial superposition between the excited levels. To verify the frequency of the observed beat signal, we first remove the exponential decay profile from the data and then fast-Fourier transform (FFT) the residual. The FFT results (see inset) confirm that the observed beat frequency is $\omega_{23}$ as expected. 

\begin{figure*}[t]
	\centering
	\includegraphics[width = 6.5 in]{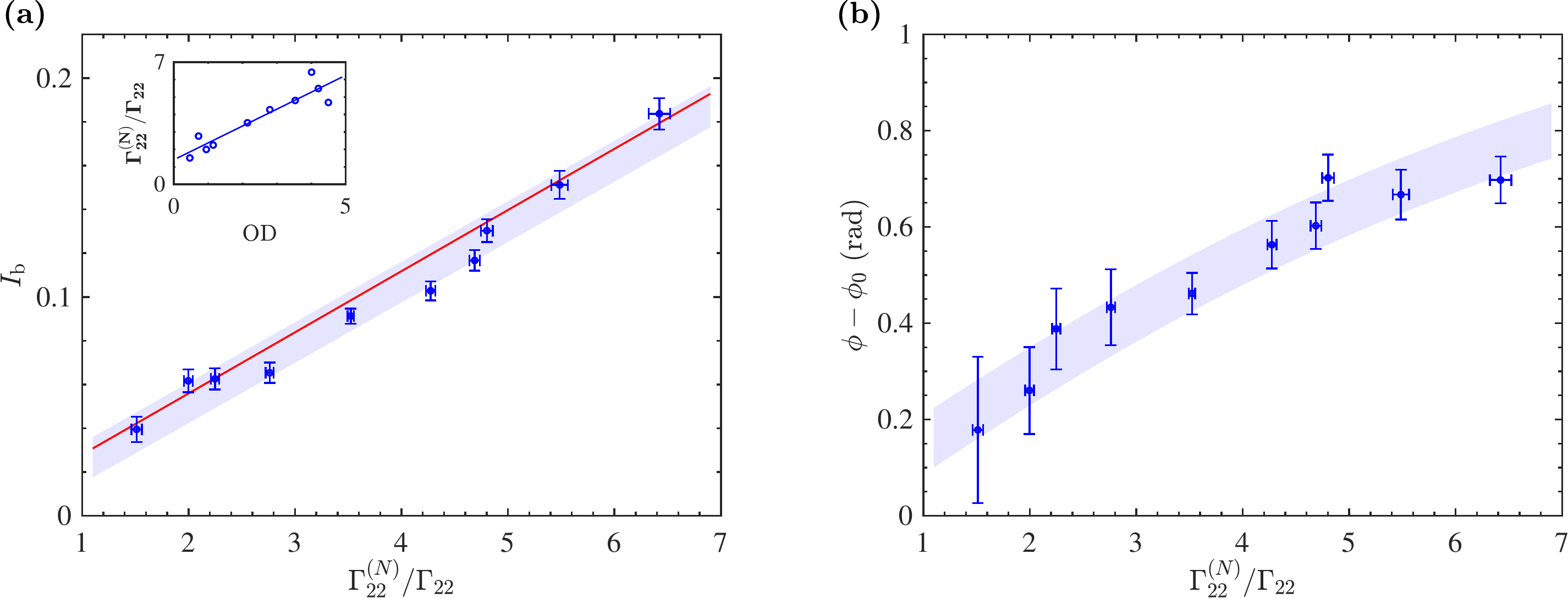}
	\caption{\textbf{(a)} The relative beat intensity $I_\mathrm{b}$ is plotted as a function of $\Gamma^{(N)}_{22}/\Gamma_{22}$ for various OD. The plotted error bars represent the one-sigma confidence interval of the fitting  to the modulated decay curves. The shaded region displays the one-sigma confidence band of a linear fit to the data. The red solid line plots Eq.\,\eqref{eq_beat_amp}. The inset shows a linear dependence of $\Gamma^{(N)}_{22}/\Gamma_{22}$ on the OD. \textbf{(b)} The beat phase $\phi$ subtracted by the common  offset $\phi_0$ is presented. The shaded region represents the one-sigma confidence band of the fitting of Eq.\,\eqref{eq_fit_phase} to the data.}
	\label{fig_phase} 
\end{figure*}

The solid lines in Fig.\,\ref{fig_decay}\,(b) fit the modulated decay curves using Eq.\,\eqref{eq_intensity} with $I_\mathrm{b}$, $\Gamma^{(N)}_{22}$, and $\phi$ as fitting parameters, and the fit results for the full range of OD between 0.5 and 4.5 are presented in Fig.\,\ref{fig_phase}. In the inset, the linear dependence of the enhancement factor $\Gamma^{(N)}_{22}/\Gamma_{22}$ on OD displays the collective nature of the emission process, in agreement with the superradiant behavior \cite{Bienaime_2011,  Bienaime_2012, Bienaime_2013,Araujo_2016, Roof_2016, Guerin_2017}. The blue solid line fitting the data provides a linear relation ${\Gamma^{(N)}_{22}}/{\Gamma_{22}} = 1.0(1)\cdot \mathrm{OD}+1.4(4)$, showing a qualitative agreement with the previous studies \footnote{See, e.g., Eq. (4) in \cite{Araujo_2016}. The previously predicted value (1/12) of the ratio between the collective decay rate enhancement and OD is different from our measured value of 1.0(1). We remark that our experimental characteristics such as the atomic density distribution or detection solid angle may have caused a deviation from the analytical prediction \cite{Bienaime_2011}.}. The relative beat intensity $I_\mathrm{b}$ is plotted as a function of $\Gamma^{(N)}_{22}/\Gamma_{22}$ in Fig.\,\ref{fig_phase}\,(a). The blue shaded region represents the one-sigma confidence band of the linear fit to the data, displaying the amplification of the quantum beat due to the increasing number of cooperative atoms. The red solid line plotting Eq.\,\eqref{eq_beat_amp} is in good agreement with the data, confirming the validity of our model.  

The measured beat phase $\phi$ is displayed in Fig.\,\ref{fig_phase}\,(b)  fit  to
\eqn{
	\phi = \arctan\bkt{\eta\cdot \frac{\Gamma^{(N)}_{22}}{\Gamma_{22}}}+\phi_0.
	\label{eq_fit_phase}
}
The fitted value of  $\phi_0=0.17$ is presumably  due to the transient intensity of the driving field during the switch-off time. From the fit, $\eta = 1.5(3)\times10^{-1}$ is almost three times larger than its expected value of $\Gamma_{22}/\omega_{23}=5.0\times 10^{-2}$ (see Eq.\,\eqref{eq_phase}). We note that non-equilibrium dynamics during the switch-off time can produce an additional OD-dependent phase delay, potentially resulting in a larger $\eta$ value than expected, which is not captured by our current model. Such an additional phase can be used to characterize the non-equilibrium dynamics of emission during the transient time, the study of which is left to future work.

\textit{Discussion.}---We have demonstrated collective quantum beats in a spontaneous emission process without an initial superposition of the excited levels in a three-level atomic system. The collective nature of the forward emission results in an enhanced coupling between the excited levels, manifested in cooperatively amplified quantum beats We observe that the enhancement factor $\Gamma^{(N)}_{22}/\Gamma_{22}$ for the collective decay rate increases with the atomic OD. The relative beat intensity also scales with $\Gamma^{(N)}_{22}/\Gamma_{22}$, 
in excellent agreement with our theoretical prediction. It signifies a combination of two different quantum interference phenomena featuring interplay between multi-level atomic structure and multi-atom collective effects which which has been the focus of many theoretical studies \cite{Agarwal_1977,  Agarwal_2001, Chow75}.

The collective enhancement of quantum beats can be a valuable tool in precision spectroscopy by enhancing beat amplitudes in systems with small signals. It can also be utilized as a source of strongly correlated photons. For example, previous works have illustrated that a system of three-level V type atoms in an interferometric setup, as in the case of a ``quantum beat laser'' \cite{Scully_1985, Scully_1987}, can exhibit strong correlations in the two-frequency emission \cite{Ohtsu_1988, Winters_1990}. It has been suggested as a means of generating or amplifying entanglement in the radiated field modes \cite{Xiong_2005, Qamar_2008}. These proposed schemes rely on the coherence between the excited atomic levels, therefore  requiring a strong classical drive to induce such coherences. Vacuum-induced collective quantum beats can circumvent the need for a classical drive, thereby avoiding additional noise, while facilitating a collective signal enhancement.

Our study of collective quantum effects can be readily combined with waveguide optics to study  interactions between distant atomic ensembles \cite{Chang_2014,Vetsch_2010,Goban_2012,Goban_2015, Yu_2014,Goban_2014,Solano_2017_review}. Recent studies have shown that such delocalized collective states can exhibit  surprisingly rich  non-Markovian dynamics \cite{Dinc_2019, Dinc_2019_PRR, Carmele_2020, Sinha_2020,Sinha_2019_proc, Calajo_2019, Sinha_2020_PRA}. A challenge in observing such exotic dynamics is that the quantum optical correlation between the multiple emitters is highly sensitive to the position of individual atoms, requiring sub-wavelength precision. Replacing the optical frequency by the beat RF frequency could allow one to bypass the strict requirements on controlling the atomic positions. An experimental investigation of  collective effects in non-Markovian regimes with multi-level atomic ensembles coupled to optical nanofibers  is within the scope of our future works \cite{Solano_2017_review}.

\textit{Acknowledgments}.---We thank Hyun Gyung Lee and Huan Q. Bui for technical support and fruitful discussions. We are also grateful to Pablo Solano and Jonathan Hoffman for helpful comments. This research is supported by the Army Research Laboratory's Maryland ARL Quantum Partnership W911NF-17-S-0003 and the Joint Quantum Institute (70NANB16H168).

\vspace{-0.2cm}

\bibliography{ref.bib}

\begin{thebibliography}{47}%
\makeatletter
\providecommand \@ifxundefined [1]{%
 \@ifx{#1\undefined}
}%
\providecommand \@ifnum [1]{%
 \ifnum #1\expandafter \@firstoftwo
 \else \expandafter \@secondoftwo
 \fi
}%
\providecommand \@ifx [1]{%
 \ifx #1\expandafter \@firstoftwo
 \else \expandafter \@secondoftwo
 \fi
}%
\providecommand \natexlab [1]{#1}%
\providecommand \enquote  [1]{``#1''}%
\providecommand \bibnamefont  [1]{#1}%
\providecommand \bibfnamefont [1]{#1}%
\providecommand \citenamefont [1]{#1}%
\providecommand \href@noop [0]{\@secondoftwo}%
\providecommand \href [0]{\begingroup \@sanitize@url \@href}%
\providecommand \@href[1]{\@@startlink{#1}\@@href}%
\providecommand \@@href[1]{\endgroup#1\@@endlink}%
\providecommand \@sanitize@url [0]{\catcode `\\12\catcode `\$12\catcode
  `\&12\catcode `\#12\catcode `\^12\catcode `\_12\catcode `\%12\relax}%
\providecommand \@@startlink[1]{}%
\providecommand \@@endlink[0]{}%
\providecommand \url  [0]{\begingroup\@sanitize@url \@url }%
\providecommand \@url [1]{\endgroup\@href {#1}{\urlprefix }}%
\providecommand \urlprefix  [0]{URL }%
\providecommand \Eprint [0]{\href }%
\providecommand \doibase [0]{http://dx.doi.org/}%
\providecommand \selectlanguage [0]{\@gobble}%
\providecommand \bibinfo  [0]{\@secondoftwo}%
\providecommand \bibfield  [0]{\@secondoftwo}%
\providecommand \translation [1]{[#1]}%
\providecommand \BibitemOpen [0]{}%
\providecommand \bibitemStop [0]{}%
\providecommand \bibitemNoStop [0]{.\EOS\space}%
\providecommand \EOS [0]{\spacefactor3000\relax}%
\providecommand \BibitemShut  [1]{\csname bibitem#1\endcsname}%
\let\auto@bib@innerbib\@empty
%</preamble>
\bibitem [{\citenamefont {Jaynes~in}(1980)}]{Jaynes_1980}%
  \BibitemOpen
  \bibfield  {author} {\bibinfo {author} {\bibfnamefont {E.~T.}\ \bibnamefont
  {Jaynes~in}},\ }\href {\doibase 10.1007/978-1-4757-0671-0_3} {\emph {\bibinfo
  {title} {Foundations of Radiation Theory and Quantum Electrodynamics}}},\
  edited by\ \bibinfo {editor} {\bibfnamefont {A.~O.}\ \bibnamefont {Barut}}\
  (\bibinfo  {publisher} {Springer US},\ \bibinfo {address} {Boston, MA},\
  \bibinfo {year} {1980})\ pp.\ \bibinfo {pages} {37--43}\BibitemShut {NoStop}%
\bibitem [{\citenamefont {Haroche}\ \emph {et~al.}(1973)\citenamefont
  {Haroche}, \citenamefont {Paisner},\ and\ \citenamefont
  {Schawlow}}]{Haroche_1973}%
  \BibitemOpen
  \bibfield  {author} {\bibinfo {author} {\bibfnamefont {S.}~\bibnamefont
  {Haroche}}, \bibinfo {author} {\bibfnamefont {J.~A.}\ \bibnamefont
  {Paisner}}, \ and\ \bibinfo {author} {\bibfnamefont {A.~L.}\ \bibnamefont
  {Schawlow}},\ }\href {\doibase 10.1103/PhysRevLett.30.948} {\bibfield
  {journal} {\bibinfo  {journal} {Phys. Rev. Lett.}\ }\textbf {\bibinfo
  {volume} {30}},\ \bibinfo {pages} {948} (\bibinfo {year} {1973})}\BibitemShut
  {NoStop}%
\bibitem [{\citenamefont {Wade}\ \emph {et~al.}(2014)\citenamefont {Wade},
  \citenamefont {\ifmmode \check{S}\else \v{S}\fi{}ibali\ifmmode~\acute{c}\else
  \'{c}\fi{}}, \citenamefont {Keaveney}, \citenamefont {Adams},\ and\
  \citenamefont {Weatherill}}]{wade_2014}%
  \BibitemOpen
  \bibfield  {author} {\bibinfo {author} {\bibfnamefont {C.~G.}\ \bibnamefont
  {Wade}}, \bibinfo {author} {\bibfnamefont {N.}~\bibnamefont {\ifmmode
  \check{S}\else \v{S}\fi{}ibali\ifmmode~\acute{c}\else \'{c}\fi{}}}, \bibinfo
  {author} {\bibfnamefont {J.}~\bibnamefont {Keaveney}}, \bibinfo {author}
  {\bibfnamefont {C.~S.}\ \bibnamefont {Adams}}, \ and\ \bibinfo {author}
  {\bibfnamefont {K.~J.}\ \bibnamefont {Weatherill}},\ }\href {\doibase
  10.1103/PhysRevA.90.033424} {\bibfield  {journal} {\bibinfo  {journal} {Phys.
  Rev. A}\ }\textbf {\bibinfo {volume} {90}},\ \bibinfo {pages} {033424}
  (\bibinfo {year} {2014})}\BibitemShut {NoStop}%
\bibitem [{\citenamefont {Hack}\ and\ \citenamefont {Huber}(1991)}]{Hack_1991}%
  \BibitemOpen
  \bibfield  {author} {\bibinfo {author} {\bibfnamefont {E.}~\bibnamefont
  {Hack}}\ and\ \bibinfo {author} {\bibfnamefont {J.~R.}\ \bibnamefont
  {Huber}},\ }\href {\doibase 10.1080/01442359109353260} {\bibfield  {journal}
  {\bibinfo  {journal} {International Reviews in Physical Chemistry}\ }\textbf
  {\bibinfo {volume} {10}},\ \bibinfo {pages} {287} (\bibinfo {year}
  {1991})}\BibitemShut {NoStop}%
\bibitem [{\citenamefont {Stolz}\ \emph {et~al.}(1991)\citenamefont {Stolz},
  \citenamefont {Langer}, \citenamefont {Schreiber}, \citenamefont
  {Permogorov},\ and\ \citenamefont {von~der Osten}}]{Stolz_1991}%
  \BibitemOpen
  \bibfield  {author} {\bibinfo {author} {\bibfnamefont {H.}~\bibnamefont
  {Stolz}}, \bibinfo {author} {\bibfnamefont {V.}~\bibnamefont {Langer}},
  \bibinfo {author} {\bibfnamefont {E.}~\bibnamefont {Schreiber}}, \bibinfo
  {author} {\bibfnamefont {S.}~\bibnamefont {Permogorov}}, \ and\ \bibinfo
  {author} {\bibfnamefont {W.}~\bibnamefont {von~der Osten}},\ }\href {\doibase
  10.1103/PhysRevLett.67.679} {\bibfield  {journal} {\bibinfo  {journal} {Phys.
  Rev. Lett.}\ }\textbf {\bibinfo {volume} {67}},\ \bibinfo {pages} {679}
  (\bibinfo {year} {1991})}\BibitemShut {NoStop}%
\bibitem [{\citenamefont {Kozin}\ \emph {et~al.}(2002)\citenamefont {Kozin},
  \citenamefont {Davydov}, \citenamefont {Ignatiev}, \citenamefont {Kavokin},
  \citenamefont {Kavokin}, \citenamefont {Malpuech}, \citenamefont {Ren},
  \citenamefont {Sugisaki}, \citenamefont {Sugou},\ and\ \citenamefont
  {Masumoto}}]{Kozin_2002}%
  \BibitemOpen
  \bibfield  {author} {\bibinfo {author} {\bibfnamefont {I.~E.}\ \bibnamefont
  {Kozin}}, \bibinfo {author} {\bibfnamefont {V.~G.}\ \bibnamefont {Davydov}},
  \bibinfo {author} {\bibfnamefont {I.~V.}\ \bibnamefont {Ignatiev}}, \bibinfo
  {author} {\bibfnamefont {A.~V.}\ \bibnamefont {Kavokin}}, \bibinfo {author}
  {\bibfnamefont {K.~V.}\ \bibnamefont {Kavokin}}, \bibinfo {author}
  {\bibfnamefont {G.}~\bibnamefont {Malpuech}}, \bibinfo {author}
  {\bibfnamefont {H.-W.}\ \bibnamefont {Ren}}, \bibinfo {author} {\bibfnamefont
  {M.}~\bibnamefont {Sugisaki}}, \bibinfo {author} {\bibfnamefont
  {S.}~\bibnamefont {Sugou}}, \ and\ \bibinfo {author} {\bibfnamefont
  {Y.}~\bibnamefont {Masumoto}},\ }\href {\doibase 10.1103/PhysRevB.65.241312}
  {\bibfield  {journal} {\bibinfo  {journal} {Phys. Rev. B}\ }\textbf {\bibinfo
  {volume} {65}},\ \bibinfo {pages} {241312} (\bibinfo {year}
  {2002})}\BibitemShut {NoStop}%
\bibitem [{\citenamefont {Bylsma}\ \emph {et~al.}(2012)\citenamefont {Bylsma},
  \citenamefont {Dey}, \citenamefont {Paul}, \citenamefont {Hoogland},
  \citenamefont {Sargent}, \citenamefont {Luther}, \citenamefont {Beard},\ and\
  \citenamefont {Karaiskaj}}]{Bylsma_2012}%
  \BibitemOpen
  \bibfield  {author} {\bibinfo {author} {\bibfnamefont {J.}~\bibnamefont
  {Bylsma}}, \bibinfo {author} {\bibfnamefont {P.}~\bibnamefont {Dey}},
  \bibinfo {author} {\bibfnamefont {J.}~\bibnamefont {Paul}}, \bibinfo {author}
  {\bibfnamefont {S.}~\bibnamefont {Hoogland}}, \bibinfo {author}
  {\bibfnamefont {E.~H.}\ \bibnamefont {Sargent}}, \bibinfo {author}
  {\bibfnamefont {J.~M.}\ \bibnamefont {Luther}}, \bibinfo {author}
  {\bibfnamefont {M.~C.}\ \bibnamefont {Beard}}, \ and\ \bibinfo {author}
  {\bibfnamefont {D.}~\bibnamefont {Karaiskaj}},\ }\href {\doibase
  10.1103/PhysRevB.86.125322} {\bibfield  {journal} {\bibinfo  {journal} {Phys.
  Rev. B}\ }\textbf {\bibinfo {volume} {86}},\ \bibinfo {pages} {125322}
  (\bibinfo {year} {2012})}\BibitemShut {NoStop}%
\bibitem [{\citenamefont {Dicke}(1954)}]{Dicke_1954}%
  \BibitemOpen
  \bibfield  {author} {\bibinfo {author} {\bibfnamefont {R.~H.}\ \bibnamefont
  {Dicke}},\ }\href {\doibase 10.1103/PhysRev.93.99} {\bibfield  {journal}
  {\bibinfo  {journal} {Phys. Rev.}\ }\textbf {\bibinfo {volume} {93}},\
  \bibinfo {pages} {99} (\bibinfo {year} {1954})}\BibitemShut {NoStop}%
\bibitem [{\citenamefont {Gross}\ and\ \citenamefont
  {Haroche}(1982)}]{Gross_1982}%
  \BibitemOpen
  \bibfield  {author} {\bibinfo {author} {\bibfnamefont {M.}~\bibnamefont
  {Gross}}\ and\ \bibinfo {author} {\bibfnamefont {S.}~\bibnamefont
  {Haroche}},\ }\href {\doibase https://doi.org/10.1016/0370-1573(82)90102-8}
  {\bibfield  {journal} {\bibinfo  {journal} {Physics Reports}\ }\textbf
  {\bibinfo {volume} {93}},\ \bibinfo {pages} {301 } (\bibinfo {year}
  {1982})}\BibitemShut {NoStop}%
\bibitem [{\citenamefont {Hegerfeldt}\ and\ \citenamefont
  {Plenio}(1993)}]{Hegerfeldt_1993}%
  \BibitemOpen
  \bibfield  {author} {\bibinfo {author} {\bibfnamefont {G.~C.}\ \bibnamefont
  {Hegerfeldt}}\ and\ \bibinfo {author} {\bibfnamefont {M.~B.}\ \bibnamefont
  {Plenio}},\ }\href {\doibase 10.1103/PhysRevA.47.2186} {\bibfield  {journal}
  {\bibinfo  {journal} {Phys. Rev. A}\ }\textbf {\bibinfo {volume} {47}},\
  \bibinfo {pages} {2186} (\bibinfo {year} {1993})}\BibitemShut {NoStop}%
\bibitem [{\citenamefont {Hegerfeldt}\ and\ \citenamefont
  {Plenio}(1994)}]{Hegerfeldt_1994}%
  \BibitemOpen
  \bibfield  {author} {\bibinfo {author} {\bibfnamefont {G.~C.}\ \bibnamefont
  {Hegerfeldt}}\ and\ \bibinfo {author} {\bibfnamefont {M.~B.}\ \bibnamefont
  {Plenio}},\ }\href {\doibase 10.1088/0954-8998/6/1/003} {\bibfield  {journal}
  {\bibinfo  {journal} {Quantum Opt.}\ }\textbf {\bibinfo {volume} {6}},\
  \bibinfo {pages} {15} (\bibinfo {year} {1994})}\BibitemShut {NoStop}%
\bibitem [{\citenamefont {Scully}\ \emph {et~al.}(2006)\citenamefont {Scully},
  \citenamefont {Fry}, \citenamefont {Ooi},\ and\ \citenamefont
  {W\'odkiewicz}}]{Scully_2006}%
  \BibitemOpen
  \bibfield  {author} {\bibinfo {author} {\bibfnamefont {M.~O.}\ \bibnamefont
  {Scully}}, \bibinfo {author} {\bibfnamefont {E.~S.}\ \bibnamefont {Fry}},
  \bibinfo {author} {\bibfnamefont {C.~H.~R.}\ \bibnamefont {Ooi}}, \ and\
  \bibinfo {author} {\bibfnamefont {K.}~\bibnamefont {W\'odkiewicz}},\ }\href
  {\doibase 10.1103/PhysRevLett.96.010501} {\bibfield  {journal} {\bibinfo
  {journal} {Phys. Rev. Lett.}\ }\textbf {\bibinfo {volume} {96}},\ \bibinfo
  {pages} {010501} (\bibinfo {year} {2006})}\BibitemShut {NoStop}%
\bibitem [{\citenamefont {Bienaim\'e}\ \emph {et~al.}(2013)\citenamefont
  {Bienaim\'e}, \citenamefont {Bachelard}, \citenamefont {Piovella},\ and\
  \citenamefont {Kaiser}}]{Bienaime_2013}%
  \BibitemOpen
  \bibfield  {author} {\bibinfo {author} {\bibfnamefont {T.}~\bibnamefont
  {Bienaim\'e}}, \bibinfo {author} {\bibfnamefont {R.}~\bibnamefont
  {Bachelard}}, \bibinfo {author} {\bibfnamefont {N.}~\bibnamefont {Piovella}},
  \ and\ \bibinfo {author} {\bibfnamefont {R.}~\bibnamefont {Kaiser}},\ }\href
  {\doibase https://doi.org/10.1002/prop.201200089} {\bibfield  {journal}
  {\bibinfo  {journal} {Fortschritte der Physik}\ }\textbf {\bibinfo {volume}
  {61}},\ \bibinfo {pages} {377} (\bibinfo {year} {2013})}\BibitemShut
  {NoStop}%
\bibitem [{\citenamefont {Bromley}\ \emph {et~al.}(2016)\citenamefont
  {Bromley}, \citenamefont {Zhu}, \citenamefont {Bishof}, \citenamefont
  {Zhang}, \citenamefont {Bothwell}, \citenamefont {Schachenmayer},
  \citenamefont {Nicholson}, \citenamefont {Kaiser}, \citenamefont {Yelin},
  \citenamefont {Lukin}, \citenamefont {Rey},\ and\ \citenamefont
  {Ye}}]{Bromley_Ye_2016}%
  \BibitemOpen
  \bibfield  {author} {\bibinfo {author} {\bibfnamefont {S.~L.}\ \bibnamefont
  {Bromley}}, \bibinfo {author} {\bibfnamefont {B.}~\bibnamefont {Zhu}},
  \bibinfo {author} {\bibfnamefont {M.}~\bibnamefont {Bishof}}, \bibinfo
  {author} {\bibfnamefont {X.}~\bibnamefont {Zhang}}, \bibinfo {author}
  {\bibfnamefont {T.}~\bibnamefont {Bothwell}}, \bibinfo {author}
  {\bibfnamefont {J.}~\bibnamefont {Schachenmayer}}, \bibinfo {author}
  {\bibfnamefont {T.~L.}\ \bibnamefont {Nicholson}}, \bibinfo {author}
  {\bibfnamefont {R.}~\bibnamefont {Kaiser}}, \bibinfo {author} {\bibfnamefont
  {S.~F.}\ \bibnamefont {Yelin}}, \bibinfo {author} {\bibfnamefont {M.~D.}\
  \bibnamefont {Lukin}}, \bibinfo {author} {\bibfnamefont {A.~M.}\ \bibnamefont
  {Rey}}, \ and\ \bibinfo {author} {\bibfnamefont {J.}~\bibnamefont {Ye}},\
  }\href {\doibase 10.1038/ncomms11039} {\bibfield  {journal} {\bibinfo
  {journal} {Nat. Commun.}\ }\textbf {\bibinfo {volume} {7}},\ \bibinfo {pages}
  {11039} (\bibinfo {year} {2016})}\BibitemShut {NoStop}%
\bibitem [{\citenamefont {Steck}()}]{Steck_Rb85}%
  \BibitemOpen
  \bibfield  {author} {\bibinfo {author} {\bibfnamefont {D.~A.}\ \bibnamefont
  {Steck}},\ }\href@noop {} {}\bibinfo {howpublished} {"Rubidium 85 D Line
  Data," available online at \url{http://steck.us/alkalidata}},\ \bibinfo
  {note} {(revision 2.2.1, 21 November 2019)}\BibitemShut {NoStop}%
\bibitem [{\citenamefont {Ara\'ujo}\ \emph {et~al.}(2016)\citenamefont
  {Ara\'ujo}, \citenamefont {Kre\ifmmode \check{s}\else
  \v{s}\fi{}i\ifmmode~\acute{c}\else \'{c}\fi{}}, \citenamefont {Kaiser},\ and\
  \citenamefont {Guerin}}]{Araujo_2016}%
  \BibitemOpen
  \bibfield  {author} {\bibinfo {author} {\bibfnamefont {M.~O.}\ \bibnamefont
  {Ara\'ujo}}, \bibinfo {author} {\bibfnamefont {I.}~\bibnamefont {Kre\ifmmode
  \check{s}\else \v{s}\fi{}i\ifmmode~\acute{c}\else \'{c}\fi{}}}, \bibinfo
  {author} {\bibfnamefont {R.}~\bibnamefont {Kaiser}}, \ and\ \bibinfo {author}
  {\bibfnamefont {W.}~\bibnamefont {Guerin}},\ }\href {\doibase
  10.1103/PhysRevLett.117.073002} {\bibfield  {journal} {\bibinfo  {journal}
  {Phys. Rev. Lett.}\ }\textbf {\bibinfo {volume} {117}},\ \bibinfo {pages}
  {073002} (\bibinfo {year} {2016})}\BibitemShut {NoStop}%
\bibitem [{\citenamefont {Chalony}\ \emph {et~al.}(2011)\citenamefont
  {Chalony}, \citenamefont {Pierrat}, \citenamefont {Delande},\ and\
  \citenamefont {Wilkowski}}]{Chalony_2011}%
  \BibitemOpen
  \bibfield  {author} {\bibinfo {author} {\bibfnamefont {M.}~\bibnamefont
  {Chalony}}, \bibinfo {author} {\bibfnamefont {R.}~\bibnamefont {Pierrat}},
  \bibinfo {author} {\bibfnamefont {D.}~\bibnamefont {Delande}}, \ and\
  \bibinfo {author} {\bibfnamefont {D.}~\bibnamefont {Wilkowski}},\ }\href
  {\doibase 10.1103/PhysRevA.84.011401} {\bibfield  {journal} {\bibinfo
  {journal} {Phys. Rev. A}\ }\textbf {\bibinfo {volume} {84}},\ \bibinfo
  {pages} {011401} (\bibinfo {year} {2011})}\BibitemShut {NoStop}%
\bibitem [{\citenamefont {Kwong}\ \emph {et~al.}(2014)\citenamefont {Kwong},
  \citenamefont {Yang}, \citenamefont {Pramod}, \citenamefont {Pandey},
  \citenamefont {Delande}, \citenamefont {Pierrat},\ and\ \citenamefont
  {Wilkowski}}]{Kwong_2014}%
  \BibitemOpen
  \bibfield  {author} {\bibinfo {author} {\bibfnamefont {C.~C.}\ \bibnamefont
  {Kwong}}, \bibinfo {author} {\bibfnamefont {T.}~\bibnamefont {Yang}},
  \bibinfo {author} {\bibfnamefont {M.~S.}\ \bibnamefont {Pramod}}, \bibinfo
  {author} {\bibfnamefont {K.}~\bibnamefont {Pandey}}, \bibinfo {author}
  {\bibfnamefont {D.}~\bibnamefont {Delande}}, \bibinfo {author} {\bibfnamefont
  {R.}~\bibnamefont {Pierrat}}, \ and\ \bibinfo {author} {\bibfnamefont
  {D.}~\bibnamefont {Wilkowski}},\ }\href {\doibase
  10.1103/PhysRevLett.113.223601} {\bibfield  {journal} {\bibinfo  {journal}
  {Phys. Rev. Lett.}\ }\textbf {\bibinfo {volume} {113}},\ \bibinfo {pages}
  {223601} (\bibinfo {year} {2014})}\BibitemShut {NoStop}%
\bibitem [{\citenamefont {Kwong}\ \emph {et~al.}(2015)\citenamefont {Kwong},
  \citenamefont {Yang}, \citenamefont {Delande}, \citenamefont {Pierrat},\ and\
  \citenamefont {Wilkowski}}]{Kwong_2015}%
  \BibitemOpen
  \bibfield  {author} {\bibinfo {author} {\bibfnamefont {C.~C.}\ \bibnamefont
  {Kwong}}, \bibinfo {author} {\bibfnamefont {T.}~\bibnamefont {Yang}},
  \bibinfo {author} {\bibfnamefont {D.}~\bibnamefont {Delande}}, \bibinfo
  {author} {\bibfnamefont {R.}~\bibnamefont {Pierrat}}, \ and\ \bibinfo
  {author} {\bibfnamefont {D.}~\bibnamefont {Wilkowski}},\ }\href {\doibase
  10.1103/PhysRevLett.115.223601} {\bibfield  {journal} {\bibinfo  {journal}
  {Phys. Rev. Lett.}\ }\textbf {\bibinfo {volume} {115}},\ \bibinfo {pages}
  {223601} (\bibinfo {year} {2015})}\BibitemShut {NoStop}%
\bibitem [{\citenamefont {Bienaim\'e}\ \emph {et~al.}(2011)\citenamefont
  {Bienaim\'e}, \citenamefont {Petruzzo}, \citenamefont {Bigerni},
  \citenamefont {Piovella},\ and\ \citenamefont {Kaiser}}]{Bienaime_2011}%
  \BibitemOpen
  \bibfield  {author} {\bibinfo {author} {\bibfnamefont {T.}~\bibnamefont
  {Bienaim\'e}}, \bibinfo {author} {\bibfnamefont {M.}~\bibnamefont
  {Petruzzo}}, \bibinfo {author} {\bibfnamefont {D.}~\bibnamefont {Bigerni}},
  \bibinfo {author} {\bibfnamefont {N.}~\bibnamefont {Piovella}}, \ and\
  \bibinfo {author} {\bibfnamefont {R.}~\bibnamefont {Kaiser}},\ }\href
  {\doibase 10.1080/09500340.2011.594911} {\bibfield  {journal} {\bibinfo
  {journal} {Journal of Modern Optics}\ }\textbf {\bibinfo {volume} {58}},\
  \bibinfo {pages} {1942} (\bibinfo {year} {2011})}\BibitemShut {NoStop}%
\bibitem [{\citenamefont {Bienaim\'e}\ \emph {et~al.}(2012)\citenamefont
  {Bienaim\'e}, \citenamefont {Piovella},\ and\ \citenamefont
  {Kaiser}}]{Bienaime_2012}%
  \BibitemOpen
  \bibfield  {author} {\bibinfo {author} {\bibfnamefont {T.}~\bibnamefont
  {Bienaim\'e}}, \bibinfo {author} {\bibfnamefont {N.}~\bibnamefont
  {Piovella}}, \ and\ \bibinfo {author} {\bibfnamefont {R.}~\bibnamefont
  {Kaiser}},\ }\href {\doibase 10.1103/PhysRevLett.108.123602} {\bibfield
  {journal} {\bibinfo  {journal} {Phys. Rev. Lett.}\ }\textbf {\bibinfo
  {volume} {108}},\ \bibinfo {pages} {123602} (\bibinfo {year}
  {2012})}\BibitemShut {NoStop}%
\bibitem [{\citenamefont {Roof}\ \emph {et~al.}(2016)\citenamefont {Roof},
  \citenamefont {Kemp}, \citenamefont {Havey},\ and\ \citenamefont
  {Sokolov}}]{Roof_2016}%
  \BibitemOpen
  \bibfield  {author} {\bibinfo {author} {\bibfnamefont {S.~J.}\ \bibnamefont
  {Roof}}, \bibinfo {author} {\bibfnamefont {K.~J.}\ \bibnamefont {Kemp}},
  \bibinfo {author} {\bibfnamefont {M.~D.}\ \bibnamefont {Havey}}, \ and\
  \bibinfo {author} {\bibfnamefont {I.~M.}\ \bibnamefont {Sokolov}},\ }\href
  {\doibase 10.1103/PhysRevLett.117.073003} {\bibfield  {journal} {\bibinfo
  {journal} {Phys. Rev. Lett.}\ }\textbf {\bibinfo {volume} {117}},\ \bibinfo
  {pages} {073003} (\bibinfo {year} {2016})}\BibitemShut {NoStop}%
\bibitem [{\citenamefont {Guerin}\ \emph {et~al.}(2017)\citenamefont {Guerin},
  \citenamefont {Rouabah},\ and\ \citenamefont {Kaiser}}]{Guerin_2017}%
  \BibitemOpen
  \bibfield  {author} {\bibinfo {author} {\bibfnamefont {W.}~\bibnamefont
  {Guerin}}, \bibinfo {author} {\bibfnamefont {M.}~\bibnamefont {Rouabah}}, \
  and\ \bibinfo {author} {\bibfnamefont {R.}~\bibnamefont {Kaiser}},\ }\href
  {\doibase 10.1080/09500340.2016.1215564} {\bibfield  {journal} {\bibinfo
  {journal} {Journal of Modern Optics}\ }\textbf {\bibinfo {volume} {64}},\
  \bibinfo {pages} {895} (\bibinfo {year} {2017})}\BibitemShut {NoStop}%
\bibitem [{Note1()}]{Note1}%
  \BibitemOpen
  \bibinfo {note} {See, e.g., Eq. (4) in \cite {Araujo_2016}. The previously
  predicted value (1/12) of the ratio between the collective decay rate
  enhancement and OD is different from our measured value of 1.0(1). We remark
  that our experimental characteristics such as the atomic density distribution
  or detection solid angle may have caused a deviation from the analytical
  prediction \cite {Bienaime_2011}.}\BibitemShut {Stop}%
\bibitem [{\citenamefont {Agarwal}(1977)}]{Agarwal_1977}%
  \BibitemOpen
  \bibfield  {author} {\bibinfo {author} {\bibfnamefont {G.~S.}\ \bibnamefont
  {Agarwal}},\ }\href {\doibase 10.1103/PhysRevA.15.2380} {\bibfield  {journal}
  {\bibinfo  {journal} {Phys. Rev. A}\ }\textbf {\bibinfo {volume} {15}},\
  \bibinfo {pages} {2380} (\bibinfo {year} {1977})}\BibitemShut {NoStop}%
\bibitem [{\citenamefont {Agarwal}\ and\ \citenamefont
  {Patnaik}(2001)}]{Agarwal_2001}%
  \BibitemOpen
  \bibfield  {author} {\bibinfo {author} {\bibfnamefont {G.~S.}\ \bibnamefont
  {Agarwal}}\ and\ \bibinfo {author} {\bibfnamefont {A.~K.}\ \bibnamefont
  {Patnaik}},\ }\href {\doibase 10.1103/PhysRevA.63.043805} {\bibfield
  {journal} {\bibinfo  {journal} {Phys. Rev. A}\ }\textbf {\bibinfo {volume}
  {63}},\ \bibinfo {pages} {043805} (\bibinfo {year} {2001})}\BibitemShut
  {NoStop}%
\bibitem [{\citenamefont {Chow}\ \emph {et~al.}(1975)\citenamefont {Chow},
  \citenamefont {Scully},\ and\ \citenamefont {Stoner}}]{Chow75}%
  \BibitemOpen
  \bibfield  {author} {\bibinfo {author} {\bibfnamefont {W.~W.}\ \bibnamefont
  {Chow}}, \bibinfo {author} {\bibfnamefont {M.~O.}\ \bibnamefont {Scully}}, \
  and\ \bibinfo {author} {\bibfnamefont {J.~O.}\ \bibnamefont {Stoner}},\
  }\href {\doibase 10.1103/PhysRevA.11.1380} {\bibfield  {journal} {\bibinfo
  {journal} {Phys. Rev. A}\ }\textbf {\bibinfo {volume} {11}},\ \bibinfo
  {pages} {1380} (\bibinfo {year} {1975})}\BibitemShut {NoStop}%
\bibitem [{\citenamefont {Scully}(1985)}]{Scully_1985}%
  \BibitemOpen
  \bibfield  {author} {\bibinfo {author} {\bibfnamefont {M.~O.}\ \bibnamefont
  {Scully}},\ }\href {\doibase 10.1103/PhysRevLett.55.2802} {\bibfield
  {journal} {\bibinfo  {journal} {Phys. Rev. Lett.}\ }\textbf {\bibinfo
  {volume} {55}},\ \bibinfo {pages} {2802} (\bibinfo {year}
  {1985})}\BibitemShut {NoStop}%
\bibitem [{\citenamefont {Scully}\ and\ \citenamefont
  {Zubairy}(1987)}]{Scully_1987}%
  \BibitemOpen
  \bibfield  {author} {\bibinfo {author} {\bibfnamefont {M.~O.}\ \bibnamefont
  {Scully}}\ and\ \bibinfo {author} {\bibfnamefont {M.~S.}\ \bibnamefont
  {Zubairy}},\ }\href {\doibase 10.1103/PhysRevA.35.752} {\bibfield  {journal}
  {\bibinfo  {journal} {Phys. Rev. A}\ }\textbf {\bibinfo {volume} {35}},\
  \bibinfo {pages} {752} (\bibinfo {year} {1987})}\BibitemShut {NoStop}%
\bibitem [{\citenamefont {Ohtsu}\ and\ \citenamefont
  {Liou}(1988)}]{Ohtsu_1988}%
  \BibitemOpen
  \bibfield  {author} {\bibinfo {author} {\bibfnamefont {M.}~\bibnamefont
  {Ohtsu}}\ and\ \bibinfo {author} {\bibfnamefont {K.}~\bibnamefont {Liou}},\
  }\href {https://doi.org/10.1063/1.99322} {\bibfield  {journal} {\bibinfo
  {journal} {Applied Physics Letters}\ }\textbf {\bibinfo {volume} {52}},\
  \bibinfo {pages} {10} (\bibinfo {year} {1988})}\BibitemShut {NoStop}%
\bibitem [{\citenamefont {Winters}\ \emph {et~al.}(1990)\citenamefont
  {Winters}, \citenamefont {Hall},\ and\ \citenamefont
  {Toschek}}]{Winters_1990}%
  \BibitemOpen
  \bibfield  {author} {\bibinfo {author} {\bibfnamefont {M.~P.}\ \bibnamefont
  {Winters}}, \bibinfo {author} {\bibfnamefont {J.~L.}\ \bibnamefont {Hall}}, \
  and\ \bibinfo {author} {\bibfnamefont {P.~E.}\ \bibnamefont {Toschek}},\
  }\href {\doibase 10.1103/PhysRevLett.65.3116} {\bibfield  {journal} {\bibinfo
   {journal} {Phys. Rev. Lett.}\ }\textbf {\bibinfo {volume} {65}},\ \bibinfo
  {pages} {3116} (\bibinfo {year} {1990})}\BibitemShut {NoStop}%
\bibitem [{\citenamefont {Xiong}\ \emph {et~al.}(2005)\citenamefont {Xiong},
  \citenamefont {Scully},\ and\ \citenamefont {Zubairy}}]{Xiong_2005}%
  \BibitemOpen
  \bibfield  {author} {\bibinfo {author} {\bibfnamefont {H.}~\bibnamefont
  {Xiong}}, \bibinfo {author} {\bibfnamefont {M.~O.}\ \bibnamefont {Scully}}, \
  and\ \bibinfo {author} {\bibfnamefont {M.~S.}\ \bibnamefont {Zubairy}},\
  }\href {\doibase 10.1103/PhysRevLett.94.023601} {\bibfield  {journal}
  {\bibinfo  {journal} {Phys. Rev. Lett.}\ }\textbf {\bibinfo {volume} {94}},\
  \bibinfo {pages} {023601} (\bibinfo {year} {2005})}\BibitemShut {NoStop}%
\bibitem [{\citenamefont {Qamar}\ \emph {et~al.}(2008)\citenamefont {Qamar},
  \citenamefont {Ghafoor}, \citenamefont {Hillery},\ and\ \citenamefont
  {Zubairy}}]{Qamar_2008}%
  \BibitemOpen
  \bibfield  {author} {\bibinfo {author} {\bibfnamefont {S.}~\bibnamefont
  {Qamar}}, \bibinfo {author} {\bibfnamefont {F.}~\bibnamefont {Ghafoor}},
  \bibinfo {author} {\bibfnamefont {M.}~\bibnamefont {Hillery}}, \ and\
  \bibinfo {author} {\bibfnamefont {M.~S.}\ \bibnamefont {Zubairy}},\ }\href
  {\doibase 10.1103/PhysRevA.77.062308} {\bibfield  {journal} {\bibinfo
  {journal} {Phys. Rev. A}\ }\textbf {\bibinfo {volume} {77}},\ \bibinfo
  {pages} {062308} (\bibinfo {year} {2008})}\BibitemShut {NoStop}%
\bibitem [{\citenamefont {Chang}\ \emph {et~al.}(2014)\citenamefont {Chang},
  \citenamefont {Vuleti{\'{c}}},\ and\ \citenamefont {Lukin}}]{Chang_2014}%
  \BibitemOpen
  \bibfield  {author} {\bibinfo {author} {\bibfnamefont {D.~E.}\ \bibnamefont
  {Chang}}, \bibinfo {author} {\bibfnamefont {V.}~\bibnamefont
  {Vuleti{\'{c}}}}, \ and\ \bibinfo {author} {\bibfnamefont {M.~D.}\
  \bibnamefont {Lukin}},\ }\href {\doibase 10.1038/nphoton.2014.192} {\bibfield
   {journal} {\bibinfo  {journal} {Nature Photonics}\ }\textbf {\bibinfo
  {volume} {8}},\ \bibinfo {pages} {685} (\bibinfo {year} {2014})}\BibitemShut
  {NoStop}%
\bibitem [{\citenamefont {Vetsch}\ \emph {et~al.}(2010)\citenamefont {Vetsch},
  \citenamefont {Reitz}, \citenamefont {Sagu\'e}, \citenamefont {Schmidt},
  \citenamefont {Dawkins},\ and\ \citenamefont {Rauschenbeutel}}]{Vetsch_2010}%
  \BibitemOpen
  \bibfield  {author} {\bibinfo {author} {\bibfnamefont {E.}~\bibnamefont
  {Vetsch}}, \bibinfo {author} {\bibfnamefont {D.}~\bibnamefont {Reitz}},
  \bibinfo {author} {\bibfnamefont {G.}~\bibnamefont {Sagu\'e}}, \bibinfo
  {author} {\bibfnamefont {R.}~\bibnamefont {Schmidt}}, \bibinfo {author}
  {\bibfnamefont {S.~T.}\ \bibnamefont {Dawkins}}, \ and\ \bibinfo {author}
  {\bibfnamefont {A.}~\bibnamefont {Rauschenbeutel}},\ }\href {\doibase
  10.1103/PhysRevLett.104.203603} {\bibfield  {journal} {\bibinfo  {journal}
  {Phys. Rev. Lett.}\ }\textbf {\bibinfo {volume} {104}},\ \bibinfo {pages}
  {203603} (\bibinfo {year} {2010})}\BibitemShut {NoStop}%
\bibitem [{\citenamefont {Goban}\ \emph {et~al.}(2012)\citenamefont {Goban},
  \citenamefont {Choi}, \citenamefont {Alton}, \citenamefont {Ding},
  \citenamefont {Lacro\^ute}, \citenamefont {Pototschnig}, \citenamefont
  {Thiele}, \citenamefont {Stern},\ and\ \citenamefont {Kimble}}]{Goban_2012}%
  \BibitemOpen
  \bibfield  {author} {\bibinfo {author} {\bibfnamefont {A.}~\bibnamefont
  {Goban}}, \bibinfo {author} {\bibfnamefont {K.~S.}\ \bibnamefont {Choi}},
  \bibinfo {author} {\bibfnamefont {D.~J.}\ \bibnamefont {Alton}}, \bibinfo
  {author} {\bibfnamefont {D.}~\bibnamefont {Ding}}, \bibinfo {author}
  {\bibfnamefont {C.}~\bibnamefont {Lacro\^ute}}, \bibinfo {author}
  {\bibfnamefont {M.}~\bibnamefont {Pototschnig}}, \bibinfo {author}
  {\bibfnamefont {T.}~\bibnamefont {Thiele}}, \bibinfo {author} {\bibfnamefont
  {N.~P.}\ \bibnamefont {Stern}}, \ and\ \bibinfo {author} {\bibfnamefont
  {H.~J.}\ \bibnamefont {Kimble}},\ }\href {\doibase
  10.1103/PhysRevLett.109.033603} {\bibfield  {journal} {\bibinfo  {journal}
  {Phys. Rev. Lett.}\ }\textbf {\bibinfo {volume} {109}},\ \bibinfo {pages}
  {033603} (\bibinfo {year} {2012})}\BibitemShut {NoStop}%
\bibitem [{\citenamefont {Goban}\ \emph {et~al.}(2015)\citenamefont {Goban},
  \citenamefont {Hung}, \citenamefont {Hood}, \citenamefont {Yu}, \citenamefont
  {Muniz}, \citenamefont {Painter},\ and\ \citenamefont {Kimble}}]{Goban_2015}%
  \BibitemOpen
  \bibfield  {author} {\bibinfo {author} {\bibfnamefont {A.}~\bibnamefont
  {Goban}}, \bibinfo {author} {\bibfnamefont {C.-L.}\ \bibnamefont {Hung}},
  \bibinfo {author} {\bibfnamefont {J.~D.}\ \bibnamefont {Hood}}, \bibinfo
  {author} {\bibfnamefont {S.-P.}\ \bibnamefont {Yu}}, \bibinfo {author}
  {\bibfnamefont {J.~A.}\ \bibnamefont {Muniz}}, \bibinfo {author}
  {\bibfnamefont {O.}~\bibnamefont {Painter}}, \ and\ \bibinfo {author}
  {\bibfnamefont {H.~J.}\ \bibnamefont {Kimble}},\ }\href {\doibase
  10.1103/PhysRevLett.115.063601} {\bibfield  {journal} {\bibinfo  {journal}
  {Phys. Rev. Lett.}\ }\textbf {\bibinfo {volume} {115}},\ \bibinfo {pages}
  {063601} (\bibinfo {year} {2015})}\BibitemShut {NoStop}%
\bibitem [{\citenamefont {Yu}\ \emph {et~al.}(2014)\citenamefont {Yu},
  \citenamefont {Hood}, \citenamefont {Muniz}, \citenamefont {Martin},
  \citenamefont {Norte}, \citenamefont {Hung}, \citenamefont {Meenehan},
  \citenamefont {Cohen}, \citenamefont {Painter},\ and\ \citenamefont
  {Kimble}}]{Yu_2014}%
  \BibitemOpen
  \bibfield  {author} {\bibinfo {author} {\bibfnamefont {S.-P.}\ \bibnamefont
  {Yu}}, \bibinfo {author} {\bibfnamefont {J.~D.}\ \bibnamefont {Hood}},
  \bibinfo {author} {\bibfnamefont {J.~A.}\ \bibnamefont {Muniz}}, \bibinfo
  {author} {\bibfnamefont {M.~J.}\ \bibnamefont {Martin}}, \bibinfo {author}
  {\bibfnamefont {R.}~\bibnamefont {Norte}}, \bibinfo {author} {\bibfnamefont
  {C.-L.}\ \bibnamefont {Hung}}, \bibinfo {author} {\bibfnamefont {S.~M.}\
  \bibnamefont {Meenehan}}, \bibinfo {author} {\bibfnamefont {J.~D.}\
  \bibnamefont {Cohen}}, \bibinfo {author} {\bibfnamefont {O.}~\bibnamefont
  {Painter}}, \ and\ \bibinfo {author} {\bibfnamefont {H.~J.}\ \bibnamefont
  {Kimble}},\ }\href {\doibase 10.1063/1.4868975} {\bibfield  {journal}
  {\bibinfo  {journal} {Applied Physics Letters}\ }\textbf {\bibinfo {volume}
  {104}},\ \bibinfo {pages} {111103} (\bibinfo {year} {2014})}\BibitemShut
  {NoStop}%
\bibitem [{\citenamefont {Goban}\ \emph {et~al.}(2014)\citenamefont {Goban},
  \citenamefont {Hung}, \citenamefont {Yu}, \citenamefont {Hood}, \citenamefont
  {Muniz}, \citenamefont {Lee}, \citenamefont {Martin}, \citenamefont
  {McClung}, \citenamefont {Choi}, \citenamefont {Chang}, \citenamefont
  {Painter},\ and\ \citenamefont {Kimble}}]{Goban_2014}%
  \BibitemOpen
  \bibfield  {author} {\bibinfo {author} {\bibfnamefont {A.}~\bibnamefont
  {Goban}}, \bibinfo {author} {\bibfnamefont {C.-L.}\ \bibnamefont {Hung}},
  \bibinfo {author} {\bibfnamefont {S.-P.}\ \bibnamefont {Yu}}, \bibinfo
  {author} {\bibfnamefont {J.~D.}\ \bibnamefont {Hood}}, \bibinfo {author}
  {\bibfnamefont {J.~A.}\ \bibnamefont {Muniz}}, \bibinfo {author}
  {\bibfnamefont {J.~H.}\ \bibnamefont {Lee}}, \bibinfo {author} {\bibfnamefont
  {M.~J.}\ \bibnamefont {Martin}}, \bibinfo {author} {\bibfnamefont {A.~C.}\
  \bibnamefont {McClung}}, \bibinfo {author} {\bibfnamefont {K.~S.}\
  \bibnamefont {Choi}}, \bibinfo {author} {\bibfnamefont {D.~E.}\ \bibnamefont
  {Chang}}, \bibinfo {author} {\bibfnamefont {O.}~\bibnamefont {Painter}}, \
  and\ \bibinfo {author} {\bibfnamefont {H.~J.}\ \bibnamefont {Kimble}},\
  }\href {\doibase 10.1038/ncomms4808} {\bibfield  {journal} {\bibinfo
  {journal} {Nature Communications}\ }\textbf {\bibinfo {volume} {5}},\
  \bibinfo {pages} {3808} (\bibinfo {year} {2014})}\BibitemShut {NoStop}%
\bibitem [{\citenamefont {Solano}\ \emph {et~al.}(2017)\citenamefont {Solano},
  \citenamefont {Grover}, \citenamefont {Hoffman}, \citenamefont {Ravets},
  \citenamefont {Fatemi}, \citenamefont {Orozco},\ and\ \citenamefont
  {Rolston}}]{Solano_2017_review}%
  \BibitemOpen
  \bibfield  {author} {\bibinfo {author} {\bibfnamefont {P.}~\bibnamefont
  {Solano}}, \bibinfo {author} {\bibfnamefont {J.~A.}\ \bibnamefont {Grover}},
  \bibinfo {author} {\bibfnamefont {J.~E.}\ \bibnamefont {Hoffman}}, \bibinfo
  {author} {\bibfnamefont {S.}~\bibnamefont {Ravets}}, \bibinfo {author}
  {\bibfnamefont {F.~K.}\ \bibnamefont {Fatemi}}, \bibinfo {author}
  {\bibfnamefont {L.~A.}\ \bibnamefont {Orozco}}, \ and\ \bibinfo {author}
  {\bibfnamefont {S.~L.}\ \bibnamefont {Rolston}},\ }\href {\doibase
  https://doi.org/10.1016/bs.aamop.2017.02.003} {\bibfield  {journal} {\bibinfo
   {journal} {Adv. At. Mol. Opt. Phys.}\ }\textbf {\bibinfo {volume} {66}},\
  \bibinfo {pages} {439 } (\bibinfo {year} {2017})}\BibitemShut {NoStop}%
\bibitem [{\citenamefont {Dinc}\ \emph {et~al.}(2019)\citenamefont {Dinc},
  \citenamefont {Ercan},\ and\ \citenamefont {Bra{\'{n}}czyk}}]{Dinc_2019}%
  \BibitemOpen
  \bibfield  {author} {\bibinfo {author} {\bibfnamefont {F.}~\bibnamefont
  {Dinc}}, \bibinfo {author} {\bibfnamefont {{\.{I}}.}~\bibnamefont {Ercan}}, \
  and\ \bibinfo {author} {\bibfnamefont {A.~M.}\ \bibnamefont
  {Bra{\'{n}}czyk}},\ }\href {\doibase 10.22331/q-2019-12-09-213} {\bibfield
  {journal} {\bibinfo  {journal} {{Quantum}}\ }\textbf {\bibinfo {volume}
  {3}},\ \bibinfo {pages} {213} (\bibinfo {year} {2019})}\BibitemShut {NoStop}%
\bibitem [{\citenamefont {Dinc}\ and\ \citenamefont {Bra\ifmmode~\acute{n}\else
  \'{n}\fi{}czyk}(2019)}]{Dinc_2019_PRR}%
  \BibitemOpen
  \bibfield  {author} {\bibinfo {author} {\bibfnamefont {F.}~\bibnamefont
  {Dinc}}\ and\ \bibinfo {author} {\bibfnamefont {A.~M.}\ \bibnamefont
  {Bra\ifmmode~\acute{n}\else \'{n}\fi{}czyk}},\ }\href {\doibase
  10.1103/PhysRevResearch.1.032042} {\bibfield  {journal} {\bibinfo  {journal}
  {Phys. Rev. Research}\ }\textbf {\bibinfo {volume} {1}},\ \bibinfo {pages}
  {032042} (\bibinfo {year} {2019})}\BibitemShut {NoStop}%
\bibitem [{\citenamefont {Carmele}\ \emph {et~al.}(2020)\citenamefont
  {Carmele}, \citenamefont {Nemet}, \citenamefont {Canela},\ and\ \citenamefont
  {Parkins}}]{Carmele_2020}%
  \BibitemOpen
  \bibfield  {author} {\bibinfo {author} {\bibfnamefont {A.}~\bibnamefont
  {Carmele}}, \bibinfo {author} {\bibfnamefont {N.}~\bibnamefont {Nemet}},
  \bibinfo {author} {\bibfnamefont {V.}~\bibnamefont {Canela}}, \ and\ \bibinfo
  {author} {\bibfnamefont {S.}~\bibnamefont {Parkins}},\ }\href {\doibase
  10.1103/PhysRevResearch.2.013238} {\bibfield  {journal} {\bibinfo  {journal}
  {Phys. Rev. Research}\ }\textbf {\bibinfo {volume} {2}},\ \bibinfo {pages}
  {013238} (\bibinfo {year} {2020})}\BibitemShut {NoStop}%
\bibitem [{\citenamefont {Sinha}\ \emph
  {et~al.}(2020{\natexlab{a}})\citenamefont {Sinha}, \citenamefont {Meystre},
  \citenamefont {Goldschmidt}, \citenamefont {Fatemi}, \citenamefont
  {Rolston},\ and\ \citenamefont {Solano}}]{Sinha_2020}%
  \BibitemOpen
  \bibfield  {author} {\bibinfo {author} {\bibfnamefont {K.}~\bibnamefont
  {Sinha}}, \bibinfo {author} {\bibfnamefont {P.}~\bibnamefont {Meystre}},
  \bibinfo {author} {\bibfnamefont {E.~A.}\ \bibnamefont {Goldschmidt}},
  \bibinfo {author} {\bibfnamefont {F.~K.}\ \bibnamefont {Fatemi}}, \bibinfo
  {author} {\bibfnamefont {S.~L.}\ \bibnamefont {Rolston}}, \ and\ \bibinfo
  {author} {\bibfnamefont {P.}~\bibnamefont {Solano}},\ }\href {\doibase
  10.1103/PhysRevLett.124.043603} {\bibfield  {journal} {\bibinfo  {journal}
  {Phys. Rev. Lett.}\ }\textbf {\bibinfo {volume} {124}},\ \bibinfo {pages}
  {043603} (\bibinfo {year} {2020}{\natexlab{a}})}\BibitemShut {NoStop}%
\bibitem [{\citenamefont {Sinha}\ \emph {et~al.}(2019)\citenamefont {Sinha},
  \citenamefont {Meystre},\ and\ \citenamefont {Solano}}]{Sinha_2019_proc}%
  \BibitemOpen
  \bibfield  {author} {\bibinfo {author} {\bibfnamefont {K.}~\bibnamefont
  {Sinha}}, \bibinfo {author} {\bibfnamefont {P.}~\bibnamefont {Meystre}}, \
  and\ \bibinfo {author} {\bibfnamefont {P.}~\bibnamefont {Solano}},\ }in\
  \href {\doibase 10.1117/12.2530927} {\emph {\bibinfo {booktitle} {Quantum
  Nanophotonic Materials, Devices, and Systems 2019}}},\ Vol.\ \bibinfo
  {volume} {11091},\ \bibinfo {editor} {edited by\ \bibinfo {editor}
  {\bibfnamefont {C.}~\bibnamefont {Soci}}, \bibinfo {editor} {\bibfnamefont
  {M.~T.}\ \bibnamefont {Sheldon}}, \ and\ \bibinfo {editor} {\bibfnamefont
  {M.}~\bibnamefont {Agio}}},\ \bibinfo {organization} {International Society
  for Optics and Photonics}\ (\bibinfo  {publisher} {SPIE},\ \bibinfo {year}
  {2019})\ pp.\ \bibinfo {pages} {53 -- 59}\BibitemShut {NoStop}%
\bibitem [{\citenamefont {Calaj\'o}\ \emph {et~al.}(2019)\citenamefont
  {Calaj\'o}, \citenamefont {Fang}, \citenamefont {Baranger},\ and\
  \citenamefont {Ciccarello}}]{Calajo_2019}%
  \BibitemOpen
  \bibfield  {author} {\bibinfo {author} {\bibfnamefont {G.}~\bibnamefont
  {Calaj\'o}}, \bibinfo {author} {\bibfnamefont {Y.-L.~L.}\ \bibnamefont
  {Fang}}, \bibinfo {author} {\bibfnamefont {H.~U.}\ \bibnamefont {Baranger}},
  \ and\ \bibinfo {author} {\bibfnamefont {F.}~\bibnamefont {Ciccarello}},\
  }\href {\doibase 10.1103/PhysRevLett.122.073601} {\bibfield  {journal}
  {\bibinfo  {journal} {Phys. Rev. Lett.}\ }\textbf {\bibinfo {volume} {122}},\
  \bibinfo {pages} {073601} (\bibinfo {year} {2019})}\BibitemShut {NoStop}%
\bibitem [{\citenamefont {Sinha}\ \emph
  {et~al.}(2020{\natexlab{b}})\citenamefont {Sinha}, \citenamefont
  {Gonz\'alez-Tudela}, \citenamefont {Lu},\ and\ \citenamefont
  {Solano}}]{Sinha_2020_PRA}%
  \BibitemOpen
  \bibfield  {author} {\bibinfo {author} {\bibfnamefont {K.}~\bibnamefont
  {Sinha}}, \bibinfo {author} {\bibfnamefont {A.}~\bibnamefont
  {Gonz\'alez-Tudela}}, \bibinfo {author} {\bibfnamefont {Y.}~\bibnamefont
  {Lu}}, \ and\ \bibinfo {author} {\bibfnamefont {P.}~\bibnamefont {Solano}},\
  }\href {\doibase 10.1103/PhysRevA.102.043718} {\bibfield  {journal} {\bibinfo
   {journal} {Phys. Rev. A}\ }\textbf {\bibinfo {volume} {102}},\ \bibinfo
  {pages} {043718} (\bibinfo {year} {2020}{\natexlab{b}})}\BibitemShut
  {NoStop}%
\end{thebibliography}%

\clearpage
\onecolumngrid
\begin{center}
	
	\newcommand{\beginsupplement}{%
		\setcounter{table}{0}
		\renewcommand{\thetable}{S\arabic{table}}%
		\setcounter{figure}{0}
		\renewcommand{\thefigure}{S\arabic{figure}}%
	}
	\beginsupplement
	
	\textbf{\large Supplemental Material}

	\title{ Supplemental Material for ``Observation of vacuum-induced collective quantum beats''}
	\author{Hyok Sang Han}
	\affiliation{Joint Quantum Institute, University of Maryland and the National Institute of Standards and Technology, College Park, Maryland 20742, USA}
	\author{Ahreum Lee}
	\affiliation{Joint Quantum Institute, University of Maryland and the National Institute of Standards and Technology, College Park, Maryland 20742, USA}
	
	\author{Kanupriya Sinha} 
	\email{kanu@princeton.edu}
	\affiliation{Department of Electrical Engineering, Princeton University, Princeton, New Jersey 08544, USA}
	
	\author{Fredrik K. Fatemi}
	\affiliation{U.S. Army Research Laboratory, Adelphi, Maryland 20783, USA}
	\author{S. L. Rolston}
	\email{rolston@umd.edu}
	\affiliation{Joint Quantum Institute, University of Maryland and the National Institute of Standards and Technology, College Park, Maryland 20742, USA}
	
\end{center}
\maketitle

\makeatletter
\renewcommand{\theequation}{S\arabic{equation}}
\renewcommand{\thefigure}{S\arabic{figure}}
\renewcommand{\bibnumfmt}[1]{[S#1]}

\section{Model}

We consider a collection of N three-level V-type atoms located at the same position. We label the ground state as $\ket{1}$ and the two excited states as $\ket{2}$ and $\ket{3}$, and the transition frequency from level j to i as $\omega_{ij}$. A weak drive field which is resonantly tuned to $\omega_{21}$ prepares the atomic system in a timed-Dicke state. As the drive field is turned off, we detect the photons emitted from the cloud in the forward direction. In the experiment, the atomic cloud has a finite size, but for theoretical simplicity we can assume it to be point-like ensemble interacting each other through the vacuum field modes. This is because we are measuring the forward scattering, where any phases of emitted photons due to the atomic position distribution is exactly compensated by the phases initially imprinted on the atoms by the drive field \cite{Scully_2006}. Additionally, the transitions $\ket{1}\leftrightarrow\ket{2}$ and $\ket{1}\leftrightarrow\ket{3}$  interact with the field effectively with the same  phase considering that the atomic cloud size is much smaller compared to $2\pi c/\omega_{23}$. We note that while the forward-scattered field is collectively enhanced, the decay rate of the atoms arising from interaction with the rest of the modes is not cooperative \cite{Bienaime_2011}.

The atomic Hamiltonian $H_A$ and the vacuum field Hamiltonian $H_F$ are
\eqn{\begin{split}
		H_A &= \sum_{m=1}^{N}\sum_{j=2,3} \hbar \omega_{j1} \hat{\sigma}_{m,j}^+ \hat{\sigma}_{m,j}^-,\\
		H_F &= \sum_{k} \hbar \omega_{k} \hat{a}_{k}^{\dagger} \hat{a}_{k},
		\label{eq:H-0}
\end{split}}
where $\hat{\sigma}_{m,j}^{\pm}$ is the raising/lowering operator acting on $m^\mr{th}$ atom and $j^\mr{th}$ level, $\hat{a}_k^{\dagger}$ and $\hat{a}_k$ are the field creation/annihilation operators of the corresponding frequency mode $\omega_{k}$, and $N$ refers to the effective number of atoms acting cooperatively in the forward direction.

First, we prepare the atomic system by a weak drive field. The atom-drive field interaction Hamiltonian is
\eqn{H_{\text{AD}}=-\sum_{m=1}^N\sum_{j=2,3} \hbar \Omega_j^m \bkt{ \hat{\sigma}_{m,j}^+ e^{-i \omega_D t} + \hat{\sigma}_{m,j}^- e^{i \omega_D t} }.\label{eq:H-AD}}
Here, $\omega_D$ is the drive frequency and $\Omega_j^m \equiv \vec{d}_{j1}^m\cdot\vec{\epsilon_{D}}\,E_D$ is the Rabi frequency of $j^\mr{th}$ level, where $\vec{d}_{j1}^{m}$ is the dipole moment of $\ket{j}\leftrightarrow\ket{1}$ transition of $m^\mr{th}$ atom, $\vec{\epsilon}_D$ is the polarization unit vector of the drive field, and $E_D$ is the electric field of the drive field. Given that the atomic ensemble is driven with the common field in our experiment, we will assume that the atomic dipoles are aligned with the drive and each other. We can thus omit the atomic labels to write $\Omega_j$.

The interaction Hamiltonian describing the atom-vacuum field interaction, under the rotating wave approximation, is given as
\eqn{
	H_{\text{AV}} = -\sum_{m=1}^N\sum_{j=2,3}\sum_{k} \hbar g_{m,j}(\omega_k) \bkt{ \hat{\sigma}_{m,j}^+\hat{a}_{k} + \hat{\sigma}_{m,j}^-\hat{a}_k^{\dagger}}.
}
Here, the atom-field coupling strength $g_{m,j}(\omega_k) \equiv \vec{d}_{j1}^{m} \cdot \vec{\epsilon}_k\sqrt{\frac{\omega_k}{2\hbar\varepsilon_0 V}}$, 
where $\vec{\epsilon}_k$ is the polarization unit vector of the field mode, $\varepsilon_0$ is the vacuum permittivity,  and $V$ is the field mode volume. As justified previously, the atomic dipoles are aligned to each other and we write $g_{j}(\omega_k)$.
Also, note that the sum over k only refers to the forward-scattered modes. The spontaneous emission arising from the rest of the modes is to be considered separately later.

\section{Driven dynamics}

\begin{figure*}[t]
	\centering
	\includegraphics[width = 3.5 in]{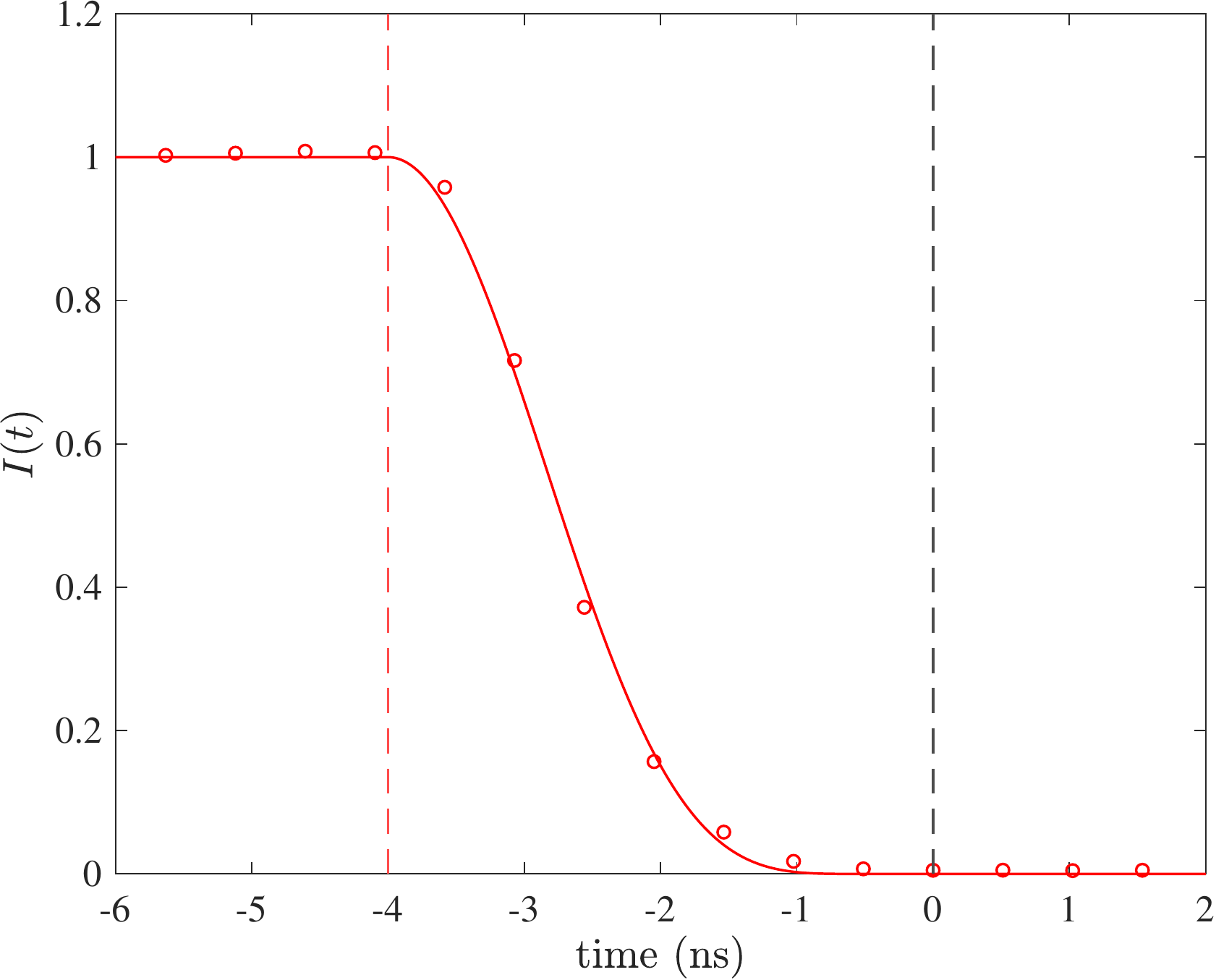}
	\caption{\textbf{(a)} The drive field intensity (red circles) at turn-off edge characterized as the truncated $\cos^4\bkt{\frac{\pi}{2}\frac{t-t_0}{\tau}}$ function (red solid line) bridging the on and off state of the intensity. Here, $t_0 = -4$ ns and the fall-time $\tau=3.5$ ns are assumed. While the intensity of the drive field turns off mostly within $\approx$ 3.5 ns, additional 0.5-ns waiting time is provided before the the data analysis of the collective emission begins at $t=0$ as shown in Fig.\,\ref{fig_decay}\,(b), to further remove the residual drive intensity and the transient effect from our measurement.}
	\label{fig_laser_extinction} 
\end{figure*}

We consider here the driven dynamics of a atoms. Moving to the rotating frame with respect to the drive frequency,  and tracing out the vacuum field modes, we can write the following Born-Markov master equation for the atomic density matrix:
\eqn{
	\der{\hat{\rho}_A}{t} = -\frac{i}{\hbar} \sbkt{\widehat{H}_A + \widehat H_{AD}, \hat {\rho}_A} - \sum_{m,n = 1}^N\sum_{i,j = 2,3} \frac{\Gamma_{ij,mn}^{(D)}}{2} \sbkt{ \hat \rho_A \widehat{\sigma}_{m,i} ^+  \widehat{\sigma}_{n,j} ^- +  \widehat{\sigma}_{m,i} ^+  \widehat{\sigma}_{n,j} ^- \hat \rho_A - 2\widehat{\sigma}_{n,j} ^- \hat \rho_A  \widehat{\sigma}_{m,i} ^+  },
}
where $\widehat H_A = - \sum_{m=1}^{N}\sum_{j=2,3} \hbar \Delta_{j} \widehat{\sigma}_{m,j}^+ \widehat{\sigma}_{m,j}^-$ is the free atomic Hamiltonian and $\widehat H_{AD} =  -\sum_{m=1}^N\sum_{j=2,3} \hbar \Omega_j^m \bkt{ \widehat{\sigma}_{m,j}^+  + \widehat{\sigma}_{m,j}^-  }$ is the atom-drive interaction Hamiltonian  in the rotating frame, with   $ \Delta_j \equiv \omega_{j1} - \omega_D$. The driven damping rates  are defined as $  \Gamma_{ij,mn}^{(D) }  \equiv \frac{\vec{d}^m_{i1} \cdot \vec{d}^n_{j1}\omega_{D}^3}{3\pi \varepsilon_0 \hbar c^3}$, with the indices $i,j$ referring to the atomic levels, and $m,n$  to different atoms.

Using the above master equation, one can obtain the following optical Bloch equations for the case of a single atom:
\begin{subequations}
	\eqn{\label{eq:optical-Bloch-eqa}
		\partial_t \rho_{33} &= i\Omega_3(\rho_{13}-\rho_{31}) - \Gamma_{33}^{(D)}\rho_{33}  - \frac{\Gamma^{(D)}_{23}}{2} \rho_{23} - \frac{\Gamma^{(D)}_{23}}{2} \rho_{32} \\
		\partial_t \rho_{22} &= i\Omega_2(\rho_{12}-\rho_{21}) - \Gamma_{22}^{(D)}\rho_{22}- \frac{\Gamma^{(D)}_{23}}{2} \rho_{23} - \frac{\Gamma^{(D)}_{23}}{2} \rho_{32}\\
		\partial_t \rho_{11} &= -i\Omega_3(\rho_{13}-\rho_{31}) -i\Omega_2(\rho_{12}-\rho_{21}) + \Gamma^{(D)}_{33}\rho_{33}+ \Gamma^{(D)}_{22}\rho_{22} +  \Gamma^{(D)}_{23}\bkt{ \rho_{23} +  \rho_{32}} \\
		\partial_t \rho_{31} &= -i \Omega_2 \rho_{32} - i \Omega_3(\rho_{33}-\rho_{11})-\bkt{\frac{\Gamma^{(D)}_{33}}{2}-i\Delta_3}\rho_{31}- \frac{\Gamma^{(D)}_{23}}{2} \rho_{21}\\
		\partial_t \rho_{13} &= i \Omega_2 \rho_{23} + i \Omega_3(\rho_{33}-\rho_{11})-\bkt{\frac{\Gamma^{(D)}_{33}}{2}+i\Delta_3}\rho_{13}- \frac{\Gamma^{(D)}_{23}}{2} \rho_{12}\\
		\partial_t \rho_{21} &= -i \Omega_3 \rho_{23} - i \Omega_2(\rho_{22}-\rho_{11})-\bkt{\frac{\Gamma^{(D)}_{22}}{2}-i\Delta_2}\rho_{21} - \frac{\Gamma^{(D)}_{23}}{2} \rho_{31}\\
		\partial_t \rho_{12} &= i \Omega_3 \rho_{32} + i \Omega_2(\rho_{22}-\rho_{11})-\bkt{\frac{\Gamma^{(D)}_{22}}{2}+i\Delta_2}\rho_{12}- \frac{\Gamma^{(D)}_{23}}{2} \rho_{13}\\
		\partial_t \rho_{32} &= -i \Omega_2 \rho_{31} + i\Omega_3\rho_{12}-\bkt{\frac{\Gamma^{(D)}_{22}+\Gamma^{(D)}_{33}}{2} -i\omega_{23}}\rho_{32} - \frac{\Gamma^{(D)}_{23}}{2} \bkt{\rho_{22} + \rho_{33}}\\
		\partial_t \rho_{23} &= i \Omega_2 \rho_{13} - i\Omega_3\rho_{21}-\bkt{\frac{\Gamma^{(D)}_{22}+\Gamma^{(D)}_{33}}{2}+i\omega_{23}}\rho_{23}  - \frac{\Gamma^{(D)}_{23}}{2} \bkt{\rho_{22} + \rho_{33}},
		\label{eq:optical-Bloch-eqi}}
\end{subequations}
where we have defined the single atom driven damping rate as $\Gamma_{ij}^{(D) }\equiv \frac{\vec{d}_{i1} \cdot \vec{d}_{j1}\omega_{D}^3}{3\pi \varepsilon_0 \hbar c^3}$.

Numerically solving Eq.\,\eqref{eq:optical-Bloch-eqa}--Eq.\,\eqref{eq:optical-Bloch-eqi} along with the normalization condition $\rho_{33}+\rho_{22}+\rho_{11}=1$ gives us the steady state density matrix $\rho_S$ for the atom. Substituting our experimental parameters, we get the populations: $\rho_{S,33}\approx 0$, $\rho_{S,22}\approx 10^{-10}$, and $\rho_{S,11}\approx 1$. The absolute value of the coherences are: $|\rho_{S,23}|\approx 0$, $|\rho_{S,21}|\approx 10^{-5}$, and $|\rho_{S,31}|\approx 0$. These estimates are made for $N \approx1 - 10$, assuming the collective driven damping rate to be $ \Gamma_{ij}^{(D)} (N) \approx (1 + Nf) \Gamma_{ij}^{(D) }$ with phenomenological value f=1 and the collective Rabi frequency to be $ \Omega_{j} \approx \sqrt{N} \Omega_{j}$. Thus we can conclude that  the atomic ensemble is well within the  single excitation regime in $\ket{2}$.

The 3.5 ns time window of laser extinction has broad spectral component and may excite extra population to $\ket{2}$ and $\ket{3}$. We numerically simulate the optical Bloch equation for this time window to find the density matrix after the laser turn-off. We model the laser turn-off shape as $\cos^4$ (see Fig. \ref{fig_laser_extinction}) and vary the Rabi frequency accordingly. Note that this is a calculation for estimate purposes and may not convey the full dynamics in the laser extinction period. Within the numerical precision limit which is set by the evolution time step ($10^{-5}$ ns) multiplied by $\Gamma_{ij} \approx 0.01$ GHz, we obtain the following density matrix values after the turn-off: $\rho_{33}\approx 0$,  $\rho_{22}\approx 0$, $\rho_{11}\approx 1$,   $\rho_{23}\approx 0$, $\rho_{12}\approx 10^{-5}$, and $\rho_{13}\approx 10^{-7}-10^{-6}$. Thus the laser turn-off edge doesn't produce any significant excitation in $\ket{3}$.

\section{Quantum beat dynamics}

As the drive field is turned off, the system evolves with the atom-vacuum field interaction Hamiltonian. Moving to the interaction representation with respect to $H_A+H_F$, we get the interaction  Hamiltonian in the interaction picture:

\eqn{
	\tilde{H}_{\text{AV}} = -\sum_{m=1}^N\sum_{j=2,3}\sum_{k} \hbar g_{j}(\omega_k) 
	\bkt{ \hat{\sigma}_{m,j}^+\hat{a}_{k}e^{i(\omega_{j1}-\omega_{k})t }
		+ \hat{\sigma}_{m,j}^-\hat{a}_k^{\dagger}e^{-i(\omega_{j1}-\omega_{k})t}},
	\label{eq:H-AV}
}

Initially the system shares one excitation in $\ket{2}$ symmetrically, and the EM field is in the vaccum state such that
\eqn{
	\ket{\Psi(0)} = \frac{1}{\sqrt{N}}\sum_{m=1}^{N}\hat{\sigma}_{m,2}^{+}\ket{11\cdots 1}\ket{\{0\}}.
	\label{eq:psi-initial}
}
As the system evolves due to the atom-vacuum field interaction, it remains in the single-excitation manifold of total atom + field Hilbert space, as one can see from the interaction Hamiltonian (Eq.\,\eqref{eq:H-AV}):
\eqn{
	\ket{\Psi(t)}=\bkt{\sum_{m=1}^{N}\sum_{j=2,3}c_{m,j}(t)\hat{\sigma}_{m,j}^{+}+\sum_{k}c_{k}(t)\hat{a}_k^{\dagger}}\ket{11\cdots 1}\ket{\{0\}}.
	\label{eq:psi-evolved}
}

Now we solve the Schr\"odinger equation to find the time evolution of the atom + field system under the atom-field interaction using Eqs.\eqref{eq:psi-evolved} and \eqref{eq:H-AV} to obtain
\begin{subequations}
	\begin{align}
	& \partial_t c_{m,j}(t) = i\sum_{k} g_j(\omega_k) e^{i(\omega_{j1}-\omega_{k})t} c_{\omega_k}(t), \\
	& \partial_t c_{\omega_k}(t) = i\sum_{m=1}^{N}\sum_{j=2,3}g_j(\omega_k)e^{-i(\omega_{j1}-\omega_{k})t}c_{m,j}(t).
	\end{align}
	\label{eq:de-1}
\end{subequations}
Formally integrating  Eq.\,\eqref{eq:de-1}(b) and plugging it in Eq.\,\eqref{eq:de-1}(a), we have
\eqn{
	\partial_t c_{m,j}(t) = -\sum_{k}g_j(\omega_k) e^{i(\omega_{j1}-\omega_{k})t}\int_0^t \dd{\tau}\sum_{n=1}^{N}\sum_{l=2,3}g_l(\omega_k) e^{-i(\omega_{l1}-\omega_k)\tau} c_{n,l}(\tau).
}
We observe that $c_{m,2}(t)$'s ($c_{m,3}(t)$'s) have the same initial conditions and the same evolution equation, thus we can justifiably define $c_2(t) \equiv c_{m,2}(t)$ ($c_3(t)\equiv c_{m,3}(t)$).

Assuming a flat spectral density of the field and making the Born-Markov approximation we get 
\begin{subequations}
	\begin{align}
	\partial_t c_2(t) &= -\frac{\Gamma_{22}^{(N)}}{2} c_2(t)-\frac{\Gamma_{23}^{(N)}}{2} e^{i\omega_{23}t}c_3(t),\\
	\partial_t c_3(t) &= -\frac{\Gamma_{33}^{(N)}}{2} c_3(t)-\frac{\Gamma_{32}^{(N)}}{2} e^{-i\omega_{23}t}c_2(t),
	\end{align}
	\label{eq:de-2}
\end{subequations}
where we have defined $\Gamma_{jl}^{(N)}\equiv \Gamma_{jl} + Nf\Gamma_{jl}$, with $\Gamma_{jl} = \frac{\vec{d}_{j1}\cdot \vec{d}_{l1}\omega_{l1}^3}{3\pi \varepsilon_0 \hbar c^3}$ as the generalized decay rate into the quasi-isotropic modes and $Nf\Gamma_{jl}$ as the collective decay rate in the forward direction \cite{Bienaime_2011, Araujo_2016}. The factor $f$ represents the geometrical factor coming from restricting the emission to the forward scattered modes. We emphasize here that the  emission into all the modes (not specifically the forward direction) denoted by $\Gamma_{jl}$ is added phenomenologically and is not collective. Considering that the atomic dipole moments induced by the drive field are oriented along the polarization of the driving field, we can obtain  $\Gamma_{23}=\sqrt{ \Gamma_{22}\Gamma_{33}}$, which can be extended to $\Gamma_{23}^{(N)}=\sqrt{ \Gamma_{22}^{(N)}\Gamma_{33}^{(N)}}$.

To solve the coupled differential equations, we take the Laplace transform of Eq.\,\eqref{eq:de-2}(a) and (b):
\begin{subequations}
	\begin{align}
	s\tilde{c}_2(s)&=c_2(0)-\frac{\Gamma_{22}^{(N)}}{2}\tilde{c}_2(s)-\frac{\Gamma_{23}^{(N)}}{2}\tilde{c}_3(s-i\omega_{23}),\\
	s\tilde{c}_3(s)&=c_3(0)-\frac{\Gamma_{33}^{(N)}}{2}\tilde{c}_3(s)-\frac{\Gamma_{32}^{(N)}}{2}\tilde{c}_2(s+i\omega_{23}),
	\end{align}
\end{subequations}
where we have defined  $\tilde{c}_j(s) \equiv \int_0^{\infty} c_j(t) e^{-st} \dd(t)$ as the Laplace transform of $c_j \bkt{t}$. Substituting  the initial conditions, we obtain the Laplace coefficients as
\begin{subequations}\begin{align}
	\tilde{c}_2(s)&=\,\frac{1}{\sqrt{N}}\frac{s+\frac{\Gamma_{33}^{(N)}}{2}-i\omega_{23}}{s^2+(\Gamma_{\text{avg}}^{(N)}-i\omega_{23})s-i\omega_{23}\frac{\Gamma_{22}^{(N)}}{2}},\\
	\tilde{c}_3(s)&=-\frac{\Gamma^{(N)}_{32}}{2\sqrt{N}}\,\frac{1}{s^2+(\Gamma^{(N)}_{\text{avg}}+i\omega_{23})s+i\omega_{23}\frac{\Gamma^{(N)}_{33}}{2}}.
	\end{align}\end{subequations}
And the poles of the denominators are, respectively,
\begin{subequations}\begin{align}
	s_{\pm}^{(2)}=&-\frac{\Gamma^{(N)}_{\text{avg}}}{2} + \frac{i\omega_{23}}{2} \pm \frac{i\delta}{2}, \\
	s_{\pm}^{(3)}=&-\frac{\Gamma^{(N)}_{\text{avg}}}{2} - \frac{i\omega_{23}}{2} \pm \frac{i\delta}{2},
	\end{align}\end{subequations}
where we have defined  $\Gamma_{\text{avg}}^{(N)}=\frac{\Gamma_{33}^{(N)}+\Gamma_{22}^{(N)}}{2}$, $\Gamma_{\text{d}}=\frac{\Gamma_{33}^{(N)}-\Gamma_{22}^{(N)}}{2}$, and $\delta = \sqrt{\omega_{23}^2-\bkt{\Gamma^{(N)}_{\text{avg}}}^2+2i\omega_{23}\Gamma^{(N)}_{\text{d}}}$. The real part of the above roots corresponds to  the collective decay rate of each of the excited states, while the imaginary part corresponds to the frequencies. The fact that $\delta$ is generally a complex number unless $\Gamma_{22}\neq\Gamma_{33}$ means that we will have modification to both the decay rate and the frequency. To see this more clearly, we can expand $\delta$ up to second order in $\Gamma_{jl}^{(N)}/\omega_{23}$, considering we are working in a spectroscopically well-separated regime ($\Gamma_{jl}^{(N)}\ll\omega_{23}$);
\eqn{
	\delta \approx \omega_{23}\sbkt{1-\frac{1}{2}\bkt{\frac{\Gamma^{(N)}_{23}}{\omega_{23}}}^2}+i\Gamma^{(N)}_{d}\sbkt{1+\frac{1}{2}\bkt{\frac{\Gamma^{(N)}_{23}}{\omega_{23}}}^2},
}
the above poles become
\begin{subequations}\begin{align}
	s_{+}^{(2)}=&-\frac{\Gamma^{(N)}_{33}}{2}\bkt{1+\frac{\Gamma^{(N)}_{\text{d}}\Gamma_{22}^{(N)}}{2\omega_{23}^2}} + i\omega_{23}\sbkt{1-\bkt{\frac{\Gamma^{(N)}_{23}}{2\omega_{23}}}^2}, \\
	s_{-}^{(2)}=&-\frac{\Gamma^{(N)}_{22}}{2}\bkt{1-\frac{\Gamma^{(N)}_{\text{d}}\Gamma_{33}^{(N)}}{2{\omega_{23}^2}}} + i\omega_{23}\bkt{\frac{\Gamma^{(N)}_{23}}{2\omega_{23}}}^2, \\
	s_{+}^{(3)}=&-\frac{\Gamma^{(N)}_{33}}{2}\bkt{1+\frac{\Gamma^{(N)}_{\text{d}}\Gamma^{(N)}_{22}}{2\omega_{23}^2}} - i\omega_{23}\bkt{\frac{\Gamma^{(N)}_{23}}{2\omega_{23}}}^2 \\
	s_{-}^{(3)}=&-\frac{\Gamma^{(N)}_{22}}{2}\bkt{1-\frac{\Gamma^{(N)}_{\text{d}}\Gamma^{(N)}_{33}}{2\omega_{23}^2}} - i\omega_{23}\sbkt{1-\bkt{\frac{\Gamma^{(N)}_{23}}{2\omega_{23}}}^2}.
	\end{align}\end{subequations}

The atomic state coefficients in time domain are
\begin{subequations}
	\begin{align}
	c_2(t) &= \frac{1}{2\sqrt{N}\delta}  e^{- \Gamma^{(N)}_{\text{avg}}t/2}  e^{i\omega_{23}t/2} \sbkt{(-i\Gamma^{(N)}_{d}-\omega_{23}+\delta)e^{i\delta t/2} + (i\Gamma^{(N)}_{d}+\omega_{23}+\delta)e^{-i\delta t/2}},\\
	c_3(t) & =\frac{i\Gamma^{(N)}_{32}}{2\sqrt{N}\delta}  e^{-\Gamma^{(N)}_{\text{avg}}t/2} e^{-i\omega_{23}t/2} \sbkt{e^{i\delta t/2}-e^{-i\delta t/2}}.
	\end{align}
	\label{eq:atom-coefficients}
\end{subequations}

Again, expanding $\delta$ under the condition $\Gamma_{jl}^{(N)}\ll\omega_{23}$, we get
\begin{subequations}
	\begin{align}
	c_2(t) &= \frac{1}{\sqrt{N}}\sbkt{ e^{- \Gamma^{(N)}_{22}t/2} - \bkt{\frac{\Gamma^{(N)}_{23}}{2\omega_{23}}}^2\frac{\delta^*}{\delta\,}e^{-\Gamma^{(N)}_{33}t/2}e^{i\omega_{23}t}},\\
	c_3(t) & = -\frac{i\Gamma^{(N)}_{32}}{2\sqrt{N}\delta} \sbkt{e^{-\Gamma^{(N)}_{22}t/2}e^{-i\omega_{23}t}-e^{-\Gamma^{(N)}_{33}t/2}}.
	\end{align}
	\label{eq:atom-coefficients-2}
\end{subequations}
Note that the collection of $N$ atoms behaves like one ``super-atom'' which decays with a  rate that is $N$-times that of an individual atom in the forward direction. We note that the system is not only superradiant with respect to the transition involving the initially excited level, but  also with respect  to other transitions as well as a result of the vacuum-induced  coupling between the levels.  Most population in $\ket{2}$ decays with the decay rate $\Gamma_{22}^{(N)}$, and small amount of it decays with $\Gamma_{33}^{(N)}$ and has corresponding level shift $\omega_{23}$. In $\ket{3}$ are the equal amount of components decaying with $\Gamma^{(N)}_{22}$ (and level shifted $-\omega_{23}$) and $\Gamma^{(N)}_{33}$. The small but nonzero contribution of $\ket{3}$ makes beating of frequency about $\omega_{23}$.

\section{Field Intensity}

The light intensity at position $x$ and time $t$ (assuming the atom is at position $x=0$ and it starts to evolve at time $t=0$) is
\eqn{
	I(x,t) = \frac{\epsilon_0 c}{2}\bra{\Psi(t)}\hat{E}^{\dagger}(x,t) \hat{E}(x,t) \ket{\Psi(t)},
}
where the electric field operator is
\eqn{
	\hat{E}(x,t) = \int_{-\infty}^{\infty} \dd k \, E_k \hat{a}_k e^{ikx}e^{-i\omega_k t}.
}
Plugging in the electric field operator and the single-excitation ansatz (Eq.\,\eqref{eq:psi-evolved}), we obtain the intensity up to a constant factor:

\eqn{
	I(x,t) \simeq N^2 \abs{e^{-i\omega_{23}\tau}c_2(\tau) + \frac{\Gamma_{23}}{\Gamma_{22}}c_3(\tau) }^2 \Theta(\tau),
	\label{eq:field-intensity}
}
where $\tau = t-\abs{x/v}$.

Substituting Eqs. \eqref{eq:atom-coefficients}(a) and (b) in the above and approximating $\delta$ in the regime $\Gamma_{jl}^{(N)}\ll\omega_{23}$, we get
\eqn{
	\frac{I(\tau)}{I_0} = e^{-\Gamma^{(N)}_{22}\tau}+\bkt{\frac{ \Gamma^{(N)}_{33}}{2\omega_{23}}}^2 e^{-\Gamma^{(N)}_{33}\tau} + \frac{\Gamma^{(N)}_{33}}{\omega_{23}} e^{-\Gamma^{(N)}_{\text{avg}}\tau} \sin(\omega_{23}\tau+\phi),
}
where $I_0$ is a normalization factor which increases as the number of atom increases. Neglecting the small second term in the right hand side, we get the relative beat intensity normalized to the main decay amplitude:
\eqn{
	\text{beat amp.} = \frac{\Gamma^{(N)}_{33}}{\omega_{23}},
}
and the beat phase $\phi$:
\eqn{
	\phi = \arctan\bkt{\frac{\Gamma^{(N)}_{22}}{\omega_{23}}}.
}
We see that even if there was no population in level 3 in the beginning, the vacuum field builds up a coherence between level 2 and level 3 to make a quantum beat. This is in line with the quantum trajectory calculation of single atom case \cite{Hegerfeldt_1994}, where the individual decay rates are replaced with collective decay rates. We can verify that the collective effect manifests in the beat size and the beat phase.

\section{Data analysis in Fig. \ref{fig_decay} (b)}

The modulated decay profiles of the flash after the peak are magnified in Fig.\,\ref{fig_decay} (b). The purpose of the figure is to visually compare the decay rate and the relative beat intensity $I_\mathrm{b}$, so we normalize each curve with the exponential decay amplitude such that the normalized intensity starts to decay from $\approx1$ at $t=0$. In practice, we fit the $I(t)$ shown in Fig.\,\ref{fig_decay} (a) after $t=0$ using Eq.\,\eqref{eq_intensity} to get $I_0$ for each curve, to get $I(t)/I_0$ curves as in Fig.\,\ref{fig_decay} (b). Note that, more precisely, it is the fitting curve that decays from $I(t)/I_0\approx1$, not the experimental data. In fact, the plotted data tend to be lower than the fitting curves near $t=0$, due to the effect of the transient behavior around the flash peak. 

The inset displays the FFT of the beat signal shown in the main figure. We first subtract from $I(t)/I_0$ data the exponential decay profile the first term of the fitting function Eq.\,\eqref{eq_intensity} as well as the dc offset. The residual, which is a sinusoidal oscillation with an exponentially decaying envelop, is the beat signal represented by the second term of Eq.\,\eqref{eq_intensity}. The FFT of the beat signal has the lower background at $\omega = 0$ due to the pre-removal of the exponential decay and the offset. The linewidth of each spectrum is limited by the finite lifetime of the beat signal, which corresponds to $\Gamma^{(N)}_\mathrm{avg}$ as in Eq.\,\eqref{eq_intensity}.

\vspace{2cm}

\end{document}